\documentclass[twocolumn]{aastex63}

\usepackage{amsmath}

\received{September 12, 2023}
\accepted{November 28, 2023}
\submitjournal{ApJ}

\begin{document}
\title{Revealing Impact of Critical Stellar Central Density on Galaxy Quenching through Cosmic Time}

\correspondingauthor{Bingxiao Xu, Yingjie Peng}
\email{bxu6@pku.edu.cn, yjpeng@pku.edu.cn}

\author{Bingxiao Xu}
 \affiliation{Department of Astronomy, School of Physics, Peking University, Yiheyuan Street No. 5, Haidian District, Beijing, China}
\affiliation{Kavli Institute of Astronomy and Astrophysics, Peking University, Yiheyuan Street No. 5, Haidian District, Beijing, China}

\author{Yingjie Peng}
\affiliation{Department of Astronomy, School of Physics, Peking University, Yiheyuan Street No. 5, Haidian District, Beijing, China}
\affiliation{Kavli Institute of Astronomy and Astrophysics, Peking University, Yiheyuan Street No. 5, Haidian District, Beijing, China}

\begin{abstract}
		In the previous work of \citet{xu21}, we investigated the 
		structural and environmental dependence on quenching in the nearby 
		universe. In this work we extend our investigations to higher 
		redshifts by combining galaxies from SDSS and ZFOURGE surveys. In 
		low density, we find a characteristic $\Sigma_{1\ kpc}$ above which
		the quenching is initiated as indicated by their 
		population-averaged color. $\Sigma^{crit}_{1\ kpc}$ shows only 
		weakly mass-dependency at all redshifts, which suggests that the 
		internal quenching process is more related to the physics that acts
		in the central region of galaxies. In high density, 
		$\Sigma^{crit}_{1\ kpc}$ for galaxies at $z > 1$ is almost 
		indistinguishable with their low-density counterparts. At $z < 1$, 
		$\Sigma^{crit}_{1\ kpc}$ for low-mass galaxies becomes 
		progressively strongly mass-dependent, which is due to the 
		increasingly stronger environmental effects at lower redshifts. 
		$\Sigma^{crit}_{1\ kpc}$ in low density shows strong redshift 
		evolution with $\sim 1$ dex decrement from $z = 2.5$ to $z = 0$. It
		is likely due to that at a given stellar mass, the host halo is on 
		average more massive and gas-rich at higher redshifts, hence a 
		higher level of integrated energy from more massive black hole is 
		required to quench. As the halo evolves from cold to hot accretion 
		phase at lower redshifts, the gas is shock-heated and becomes more 
		vulnerable to AGN feedback processes, as predicted by theory. 
		Meanwhile, angular momentum quenching also becomes more effective 
		at low redshifts, which complements a lower level of integrated 
		energy from black hole to quench.
\end{abstract}


\section{Introduction} \label{sec:intro}
One of the long-standing puzzles of galaxy evolution is to understand how 
and why the star-forming activity in galaxies is seized (``quenched''). The
processes to quench the star formation can be broadly classified into two
categories \citep{kau03b,bal06,pen10}: internally driven processes, named 
as ``mass quenching" that operates in both central and satellite galaxies;
and externally driven processes, known as ``environment quenching" which 
only operates in satellite galaxies. Various mechanisms have been proposed 
to account for the underlying physics. For mass quenching, the candidate 
mechanisms include AGN feedback \citep{cro06,dar15,dar16,lin16,del19}, 
morphological quenching \citep{mar09}, gravitational quenching 
\citep{gen14}, and angular momentum quenching \citep{pen20,ren20}. 
Mechanisms for environment quenching consist of strangulation \citep{bal97}
, ram-pressure stripping \citep{gun72,aba99}, tidal interaction 
\citep{sob11}, and halo quenching \citep{dek06}. Despite equipped with 
various options, a definitive concensus on which processes contribute to 
what extent is still lacking.

Attempts to push our understanding of quenching forward have been 
extensively made in investigating the correlations between various physical
parameters and the quiescence of galaxies. Studies of massive galaxies 
suggest that the surface mass density within the central radius of 
1 kpc ($\Sigma_{1\ kpc}$) is strongly correlated with the quenched fraction
of galaxies, hence can be treated as an effective probe to quenching. The 
usage of $\Sigma_{1\ kpc}$ was first reported in \citet{che12}, who found 
that high $\Sigma_{1\ kpc}$ performs best in predicting quenching at 
$z \sim 0.7$. \citet{fan13} found that for nearby Sloan Digital Sky Survey
(SDSS) galaxies, specific star formation rate (sSFR) varies systematically 
relative to $\Sigma_{1\ kpc}$, suggesting a mass-dependent threshold of 
$\Sigma_{1\ kpc}$ for the onset of quenching, possibly due to a threshold 
in black hole mass. \citet{van14,tac15,bar17} extended the use of 
$\Sigma_{1\ kpc}$ as a predictor of quenching to galaxies at higher 
redshifts. \citet{whi17} studied the population-averaged sSFR as a function
of $\Sigma_{1\ kpc}$ for galaxies at $0.5 < z < 2.5$, and found a sharp 
decrease in sSFR as $\Sigma_{1\ kpc}$ exceeds some threshold. They also 
found that the critical $\Sigma_{1\ kpc}$ has strong redshift evolution. 
\citet{che20} proposed an analytic model to explain the quenching 
boundaries as a competition of halo binding energy with the integrate power
of AGN feedback. \citet{luo20} found that the offset to the running median 
of $\Sigma_{1\ kpc}$ has the power of distinguishing the bulge types in 
nearby galaxies. More recently, \citet{xu21} use a sample of nearby SDSS 
galaxies to study the distribution of population-averaged (NUV - r) color 
on the $M_{\star}-\Sigma_{1\ kpc}$ plane, and its environmental dependence.
They found that for central galaxies in low density, there exists a 
critical central density $\rm \Sigma^{crit}_{1\ kpc} \sim 10^9-10^{9.2} 
M_{\odot}\,kpc^{-2}$, above which the quenching initiates. Intriguingly, 
this $\Sigma^{crit}_{1\ kpc}$ is only weakly dependent on the stellar mass.

Surprisingly, $\Sigma_{1\ kpc}$ appears also correlated with the quiescence
of satellite galaxies. \citet{woo17} showed that $\Sigma_{1\ kpc}$ in 
quenched satellites is $\sim 0.3$ dex higher than that of star-forming 
satellites at fixed stellar mass. \citet{kaw17,guo21} reach similar 
conclusions for satellites at high redshifts. \citet{xu21} find that the 
critical $\Sigma_{1\ kpc}$ at the transition from star-forming to 
passive populations is strongly mass-dependent for low-mass satellites. 
Moreover, they found that the mass-dependence in $\Sigma^{crit}_{1\ kpc}$ 
for low-mass satellites is a function of environment: 
$\Sigma^{crit}_{1\ kpc}$ is lower in dense environment, at fixed stellar 
mass.

To gain further insight of the underlying physics of quenching, a natural 
logic is to extend the work of \citet{xu21} to higher redshifts, to study 
the redshift evolution of the structural and environmental impacts on 
quenching, which is the goal of this paper. In this work, we utilize the 
samples of galaxies from Sloan Digital Sky Survey (SDSS) and The FourStar
Galaxy Evolution (ZFOURGE) surveys to perform a joint 
analysis on the structural and environmental dependence on quenching at 
$0 < z < 2.5$. Photometric redshift based on broad-band photometry 
with large uncertainty is the main obstacle to study the galaxy environment
at high redshifts. ZFOURGE survey utilize five near-IR medium-band filters 
to better constrain the photometric redshift, which enables more precise 
characterization of the environment at high redshifts. Throughout, we adopt
the following cosmological parameters where appropriate: $H_0$ = 70 km 
$s^{-1}$Mpc$^{-1}$, $\Omega_m$ = 0.3, and $\Omega_{\lambda}$ = 0.7.

\section{Data and Analysis} \label{sec:data}
\subsection{The nearby galaxies}
In this work, we use the same sample of nearby galaxy as used in 
\citet{xu21}, which was constructed from Sloan Digital Sky Survey (SDSS) 
DR7 catalog \citep{aba09}. The redshift range is $0.02 < z < 0.085$, which 
guarantees reliable spectroscopic redshift measurements. Each galaxy is 
weighted by 1/TSR $\times$ 1/V$_{max}$ , where TSR is the spatial target 
sampling rate, determined using the fraction of objects that have spectra 
in the parent photometric sample within the minimum SDSS fiber spacing of 
55" of a given object. The V$_{max}$ values are derived from the 
$k$-correction program version 4.2 \citep{bla07}. The use of V$_{max}$ 
weighting allows us to correct the effect of incompleteness of the sample 
down to a stellar mass of about $10^9\ M_{\odot}$.

Integrated photometries in five bands were used in this study: 
$u, g, r, i, z$ bands from SDSS. The photometries were corrected for 
Galactic extinction and $k$-weighted to $z = 0$ using version 4.2 of the 
$k$-correct code package described in \citet{bla07}. The spectroscopic 
redshifts, total stellar mass, fiber velocity dispersion, and median 
signal-to-noise ratios (S/Ns) in the spectra were obtained from the MPA/JHU
DR7 value-added catalog. The stellar masses were computed by fitting the 
integrated SDSS photometry with the stellar population models (similar to 
the method in \citet{sal07}). The structural parameters such as Sersic 
index $n$, effective radius $R_e$, ellipticity $e$ are obtained from 
\citep{sim11}. The axis ratio is computed as $b/a = 1 - e$ as defined. 

\subsection{The galaxies at high redshift}
We select galaxies at $0.5 < z < 2.5$ from The FourStar Galaxy Evolution
(ZFOURGE) survey \citep{str16}. The survey is composed of three 11' 
$\times$ 11' fields with coverage in the regions of CDFS \citep{gia02}, 
COSMOS \citep{sco07}, and UDS \citep{law07} that overlap with the Cosmic 
Assembly Near-IR Deep Extragalactic Legacy Survey (CANDELS; 
\citet{gro11,koe11}), which also provide Hubble Space Telescope (HST), 
high-angular resolution imaging for 0.6 - 1.6 $\mu$m (see, e.g., 
\citet{van12}). ZFOURGE utilize five near-IR medium-bandwidth filter: 
$J_1$, $J_2$, $J_3$, $H_s$ and $H_l$ to better constrain the photometric 
redshift. The medium-band near-IR imaging in $J_1$, $J_2$, $J_3$ reaches 
depths of $\sim 26$ AB mag and $\sim$ 25 AB mag in  $H_s$ and $H_l$. We 
utilize ZFOURGE main catalogs which are provided by the official ZFOURGE 
website\footnote{https://zfourge.tamu.edu/data/}. The main catalogs are 
complete for galaxies to $K_s \sim 25.5-26.0$ AB mag (see \citet{str16}), 
and include photometric redshifts and rest-frame colors calculated using 
EAZY \citep{bra08} from 0.3 to 8 $\mu$m photometry for each galaxy. The 
typical photometric-redshift uncertainties are $\sigma_z / ( 1 + z )$ = 
0.01-0.02 to the $K_s$-band magnitude limit for galaxies between $z = 0.5$ 
and $z = 2.0$, with negligible dependence on galaxy color \citep{str16}. In
addition, the morphological data that are cross-matched with Hubble Space 
Telescope (HST)/WFC3/F160W CANDELS data from \citet{van12} are also 
included.

\subsection{Sample selections}
We select the nearby galaxies above the SDSS spectroscopic limit ($r = 
17.77$) and with the stellar mass log$(M_{\star}/M_{\odot}) > 9$. We 
discard galaxies with low axis ratio with $b/a < 0.5$ to minimize the 
measurement bias due to the internal dust extinction. A final sample of 
89,469 nearby galaxies is produced for the subsequent analysis. 

For galaxies at high redshifts, we first select all the well-detected 
galaxies (`USE' flag = 1) with the stellar mass log$(M_{\star}/M_{\odot}) 
> 9$; we then discard galaxies with $H$-band magnitude fainter than 24.5 AB
mag to guarantee an accurate structural measurement \citep{van12}. To see 
if this additional magnitude cut of $H < 24.5$ AB mag has any impact
on the original stellar mass completeness determined based on the limit of 
$K_s \sim 25.5 - 26.0$ AB mag \citep{kaw17}, we compute the fraction of 
galaxies that have both $H < 24.5$ and $K_s < 25.5$ to the galaxies that 
have $K_s < 25.5$ only to evaluate the impact of the cut $H < 24.5$ (see 
Table \ref{tab:frac}). The fraction is higher than $97 \%$ at all 
redshifts, which indicates that the additional cut on H-band will not 
affect the level of completeness of the sample.
\begin{deluxetable}{ccc}
		\label{tab:frac}
		\tablehead{		
		\colhead{$z$} & \colhead{} & 
		\colhead{$\frac{N_g (H < 24.5 \,\&\, K_s < 25.5)}{N_g (K_s < 25.5)}$}}
	\startdata
		$0.5 < z < 1.0$ & \qquad    & 99.31\% \\
		$1.0 < z < 1.5$ & \qquad    & 99.10\% \\
		$1.5 < z < 2.0$ & \qquad    & 99.45\% \\
		$2.0 < z < 2.5$ & \qquad    & 97.45\% \\
	\enddata
\caption{Fraction of galaxies that satisfy both $H < 24.5$ and $K_s < 25.5$
		to the galaxies that have $K_s < 25.5$ only, at four redshift bins.}
\end{deluxetable}		

Similarly, we discard galaxies from ZFOURGE catalog with axis ratio $b/a < 
0.5$ to ensure a reliable measurement of Sersic index $n$ and $R_e$. In 
addition, we use the quality flag to further exclude galaxies with bad 
GALFIT fitting in F160W band (flag $>$ 1). A final sample of 4,577 galaxies
at $0.5 < z < 2.5$ makes the cut.

\subsection{The Central 1kpc Mass Density}
We follow the procedures in \citet{xu21} to compute the central 1kpc 
surface mass density $\Sigma_{\rm 1\ kpc}$, by directly integrating the 
Sersic light profile and scaling the integrated luminosity within the inner
1 kpc. This method has been widely used in many previous studies 
\citep{bez09,whi17,kaw17} and is described as follows. The two-dimensional 
Sersic light profile can be described in the form of
  \begin{equation} \label{eqn:sersic}
        I(r) = I_0 {\rm exp}\left[-b_n\left(\frac{r}{r_e}\right)^{1/n}\right],
  \end{equation}

where $I_0$ is the central intensity, $n$ is the Sersic indices, $r_{\rm 
eff}$ is the circularized effective radii, and $b_n$ is defined as \citep{cio99}:
  \begin{equation}
    b_n \approx 2n - \frac{1}{3} + \frac{4}{405n} + \frac{46}{25515n^2}.
  \end{equation}

For the disk galaxies with Sersic indices $n < 2.5$ \citep{ken15}, the total
luminosity is obtained by integrating over the two-dimensional light profile
(Equation \ref{eqn:sersic}). We then convert the total luminosity to the
total stellar mass, assuming that the mass follows the light and that there
are no strong color gradients. Finally, we calculate the stellar mass
surface density in the inner 1 kpc by numerically integrating the following
equation:

  \begin{equation}
		\Sigma_{\rm 1\ kpc} = \frac{\int_0^{\rm 1\ kpc}I(r)rdr }{\int_0^{\infty}I(r)rdr} \frac{L_{model}}{L_{phot}}  \frac{M_{\star}}{\pi(\rm 1\ kpc)^2}, \quad  n < 2.5
  \end{equation}
where $\rm M_{\star}$ is the total stellar mass of the galaxy from the
MPA/JHU DR7 value-added catalog for nearby galaxies, and from the ZFOURGE 
main catalogs for galaxies at high redshifts; $L_{model}$ is the total 
luminosity from the Sersic modeling, whereas $L_{phot}$ is the measured 
total luminosity within the aperture. There is only slight difference 
$\sim$ 0.1 dex between $L_{model}$ and $L_{phot} $\citep{whi17,kaw17}, and 
we do not include this correction and set $L_{model}/L_{phot} = 1$ in this
study, to maintain consistency of the computed $\Sigma_{\rm 1\ kpc}$ in 
nearby and distant galaxies. For galaxies that have prominent bulge 
components with $n > 2.5$, we assume that they follow spherical light 
profiles and perform an Abel transform to deproject the circularized, 
three-dimensional light profile \citep{bez09}:

  \begin{equation}  \label{eqn:abel}
\rho\left(\frac{r}{r_e}\right) 
         =  \frac{b_n}{\pi}\frac{I_0}{r_e}
             \left(\frac{r}{r_e}\right)^{1/n-1} \times \int_1^{\infty} \frac{{\rm exp}[-b_n(r/r_e)^{1/n}t]}{\sqrt{t^{2n} - 1}}dt.
  \end{equation}
The total luminosity in this case is derived by integrating over the above
three-dimensional light profile, and the central surface mass density is
given as
  \begin{equation}
  \Sigma_{\rm 1\ kpc} = \frac{\int_0^{\rm 1\ kpc}\rho(r)r^2dr }{\int_0^{\infty}\rho(r)r^2dr}\frac{M_{\star}}{\pi(\rm 1\ kpc)^2}, \quad  n > 2.5. 
  \end{equation}

For each galaxy, we perturb the stellar mass $M_{\star}$, Sersic index $n$ 
and size $R_e$ within their quoted 1$\sigma$ error for 40 times and compute
the corresponding $\Sigma_{1\ kpc}$. The uncertainty of $\Sigma_{1\ kpc}$ 
is evaluated as the standard deviation of the 40 perturbed $\Sigma_{1\ kpc}$.

\subsection{Characterization of Galaxy Environment}
Most methods to define and compute the environment of galaxies fall into 
two categories: those that use flexible apertures whose size is determined 
by the method of nearest neighbour, and those that use fixed apertures. The
choice is largely dependent on the scale being probed: the local 
environment that is internal to a halo is found to be best measured with 
the nearest neighbour method; whereas the fixed apertures best quantify the
large-scale environment external to a halo \citep{mul12}. In this study, we
characterize the environment of galaxies in local and distant universe by 
their local projected overdensity using the distance to the $N$th nearest 
neighbour. The dimensionless overdensity 1 + $\delta$ is defined as 
\citet{pen12}:
  \begin{equation}
  (1 + \delta)_5 = 1 + \frac{\Sigma_5 - \langle\Sigma\rangle}
               {\langle\Sigma\rangle}.
  \end{equation}

Since there is no physical constraints on the number $N$ yet, the choice of
$N$ typically varies from 3 to 10, which largely depends on the surveys 
\citep{mul12,pen12,kaw17}. For nearby galaxies from SDSS, we adopt $N = 5$ 
and the overdensity is computed from the volume of the cylinder that 
centered on each galaxy with a length $\pm$1000 $\rm km s^{-1}$. All the 
five closest neighbor galaxies have $M_{B,AB} \le -19.3 - z$, where $-z$ is
used to approximate the luminosity evolution of both passive and active 
galaxies. For galaxies at high redshift, since the sample with available
spectroscopic redshift is very limited, the photometric redshift with 
larger uncertainty is used to characterize the environment at high 
redshifts. We adopt an empirical approach to optimize $N$ and the redshift 
interval $\delta z$ (or the length of the cylinder that centered on each 
galaxy), which are vital to determine the overdensity. We use $N = 8$ and 
$\delta z = 0.08$ in this study. The detail of the precedures can be found 
in Appendix \ref{appen:otest}. 

\subsection{Star-forming Indicator}
We use (U - V) color as the indicator of star formation in this study, as 
is widely used in literatures. The rest-frame flux in U and V bands were 
computed by SED fitting and provided in ZFOURGE ``REST-FRAME" catalogs. For
nearby SDSS galaxies, \citet{bla07} provided sets of empirical formula in 
their Table 2 to convert $u, g, r$ photometries to U and V magnitudes, 
which is given as
  \begin{equation}
    \begin{aligned}		
	  U  = u - 0.0682 - 0.0140[(u - g) - 1.2638] \\ 
	  V  = g - 0.3516 - 0.7585[(g - r) - 0.6102].
    \end{aligned}	\label{eqn:color}	  
  \end{equation}		

We use Eqn \ref{eqn:color} to convert the SDSS photometries to rest-frame 
(U - V) color to be in line with the high-redshift galaxies. The provided 
color dispersion of $\sigma[u - g] = 0.26$ and $\sigma[g - r] = 0.15$ were 
used to estimated the uncertainty of the converted (U - V) color.

\subsection{Dust Extinction Correction} \label{subsec:cor}
Massive dusty galaxies with intense star-forming activity at high redshifts
typically show red color, which makes them indistinguishable with passive
galaxies based only on (U - V) color. In literatures, UVJ diagram is widely
used to effectively break this color degeneracy \citep{wil10}, and classify
galaxies as star-forming galaxies (SFGs) and quiescent galaxies (QGs). 
However, a continuous measurement of level of star formation, instead of a 
dichotomy of galaxies suits this study more. Therefore, we attempt to 
assume a Calzetti law to correct the rest-frame (U - V) color for the 
effect of dust extinction. For galaxies at high redshifts, $A_V$ is 
computed from SED fitting and is provided in the ZFOURGE main catalogs. The
value of $R_V = A_V / E(B - V)$ depends on the interstellar environment 
along the line of sight. In galactic diffuse regions, $R_V$ typically has 
an average value of 3.1 \citep{dra03}; whereas in dense molecular clouds, 
$R_V$ could be as large as $\sim 6$ \citep{mat90,fit99}, and it could be 
as $\sim 2$ in low density region \citep{fit99}. A detailed evaluation of 
$R_V$ for different types of galaxies in our sample is definitely beyond the scope of this paper. Instead, we adopt an empirical methodology to 
``optimize" the value of $R_V$ to be in line with the classification based 
on the UVJ diagram. Overall, the classification based on the corrected 
(U - V) color best matches that on UVJ diagram when $R_V \sim 5.1$ (see 
details in Appendix \ref{appen:dust}), and we adopt this value of $R_V$ to 
correct the (U - V) color. For nearby SDSS galaxies, $A_V$ for each galaxy 
is obtained by corss-matching our SDSS sample with the Galaxy Evolution 
Explorer (GALEX)-SDSS-Wide-field Infrared Survey Explorer (WISE) LEGACY 
CATALOG \citep[GSWLC,][]{sal07}. To maintain consistency with galaxies at 
high redshifts, we use the same value of $R_V \sim 5.1$ to correct the 
(U - V) color. 

\begin{figure*}
      \plotone{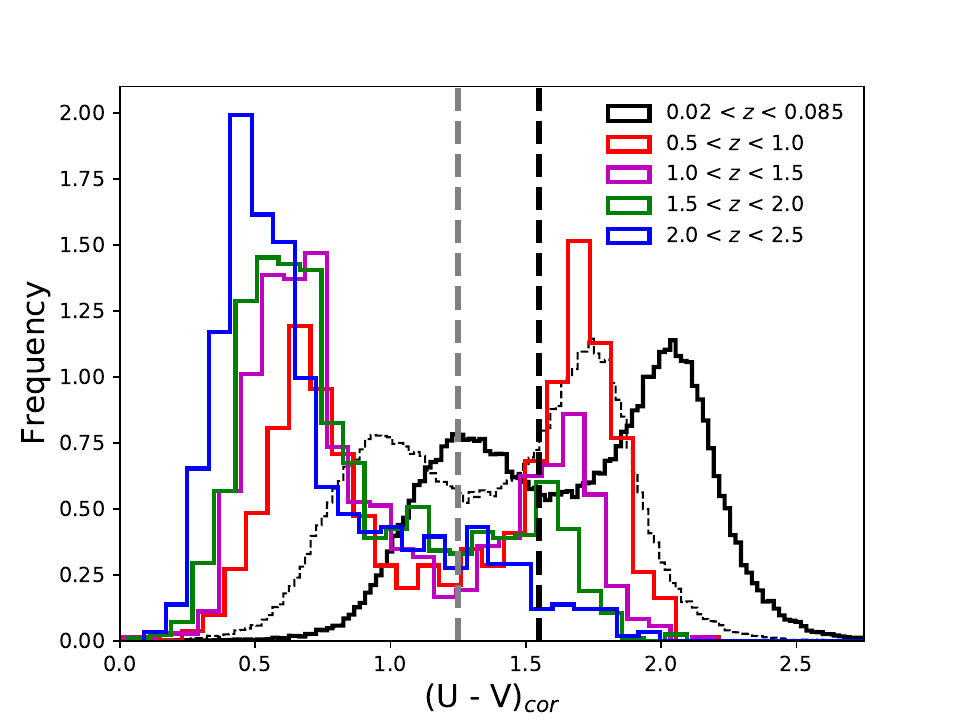}
       \caption{Distribution of dust-extinction corrected rest-frame (U - V)
		color at five redshift bins. Galaxies are selected with stellar mass
		$\rm log(M_{\star} / M_{\odot}) > 9.8$ to guarantee the completeness
		up to $z \sim 2.5$ \citep{kaw17}. Color bimodality can be observed 
		at almost all the redshit bins except for the highest one at 
		$2.0 < z < 2.5$. The gray dashed line marks the color criteria that 
		separates the star-forming and passive populations for galaxies at 
		high redshifts, which is (U - V)$_{cor} \sim 1.25$. This
		color criteria is insensitive to the redshift and the level of 
		completeness of the sample. The black dashed line marks the position
		of color trough that separates two peaks for SDSS galaxies, which 
		is at $\sim$ 1.55. To better visualize the comparison between 
		nearby and distant galaxies, we narrow down the color range of the
		whole sample by shifting the color distribution of 
		SDSS galaxies by 0.3 dex towards the left (thin dashed line), to 
		align the color trough of SDSS galaxies with those of galaxies at 
		high redshifts. This shifting will not affect the determination of 
		$\Sigma^{crit}_{1\ kpc}$, since the color criteria used in 
		computing $\Sigma^{crit}_{1\ kpc}$ shifts for the same amount as 
		for the whole distribution.
		\label{fig:uvc}}
\end{figure*}

Figure \ref{fig:uvc} shows the comparison of the distribution of the 
dust-extinction corrected (U - V) color at $0 < z < 2.5$. Only galaxies 
with $\rm log(M_{\star} / M_{\odot}) > 9.8$ are selected to maintain the 
completeness level of the sample up to $z \sim 2.5$. Overall, the color of 
galaxies becomes redder as the redshift decreases. The color bimodality can
be clearly observed at almost all the redshifts except for highest redshift
bin at $2 < z < 2.5$. For galaxies at $z > 0.5$, the postion of the trough 
between two peaks is (U - V)$_{cor}$ $\sim$ 1.25, and does not show 
significant redshift evolution. The color at $z \sim 0$ is $\sim$ 0.3 - 0.7
dex redder than those at $z > 0.5$, which is consistent with the previous 
result in which the rest-frame (U - V) color was directly derived from SED 
fitting \citep{bel12}. To better visualize the color distribution in 
galaxies at high redshifts and compare with that of the nearby galaxies, we
narrow down the color range of the whole sample by shifting the color 
distribution of SDSS galaxies by $\sim$ 0.3 dex towards the left in Figure 
\ref{fig:uvc} (black dashed line), to align the color trough of SDSS 
galaxies with those of galaxies at high redshifts. This shifting will not 
affect the subsequent determination of the critical $\Sigma_{1\ kpc}$, 
which only depends on the relative position of the trough within the 
distribution. 

In addition, we test if this color criteria is sensitive to the level of 
completeness of the sample. We repeatedly adjust the lower bound of stellar
mass of the sample and re-plot the color distribution, and found the 
location of the trough remains similar. Therefore, we use (U - V)$_{cor} 
\sim 1.25$ as a color criteria for the subsequent analysis.

\section{The Structural and Environmental impact on quenching} \label{sec:result}
In this section, we study the structural and environmental impacts on 
quenching for galaxies at $0 < z < 2.5$ by investigating the color 
distribution on the $\rm M_{\star}-\Sigma_{1\ kpc}$ plane. We assign 
galaxies into five redshift bins to study their redshift evolution. To 
reveal their environmental dependence, we divide the galaxies in each 
redshift bin into three envrionment bins based on their rank in local 
overdensity\footnote{Due to the relatively large uncertainty in the derived
photometric redshift, it is challenging to accurately identify the galaxy
clusters or groups at high redshifts. We therefore do not apply the 
``central" and ``satellite" dichotomy, but only use the local overdensity 
to characterize the environment of nearby galaxies to maintain consistentcy
with the definition of environment at high redshifts.}. For each SDSS 
galaxy, we perform a $V_{max}$-weighting correction to correct for the 
incompleteness, inside a box of 0.3$\times$0.2 dex$^2$ that centers on each
data point. We further smooth the data using the locally weighted 
regression method LOESS \citep{cle88} as implemented by \citet{cap13}. 
LOESS is useful in unveiling the overall underlying trends by reducing the 
intrinsic and observational errors, in particular in bins where the number 
of galaxies is small.

\begin{figure*}
	\epsscale{1.2}	
		\plotone{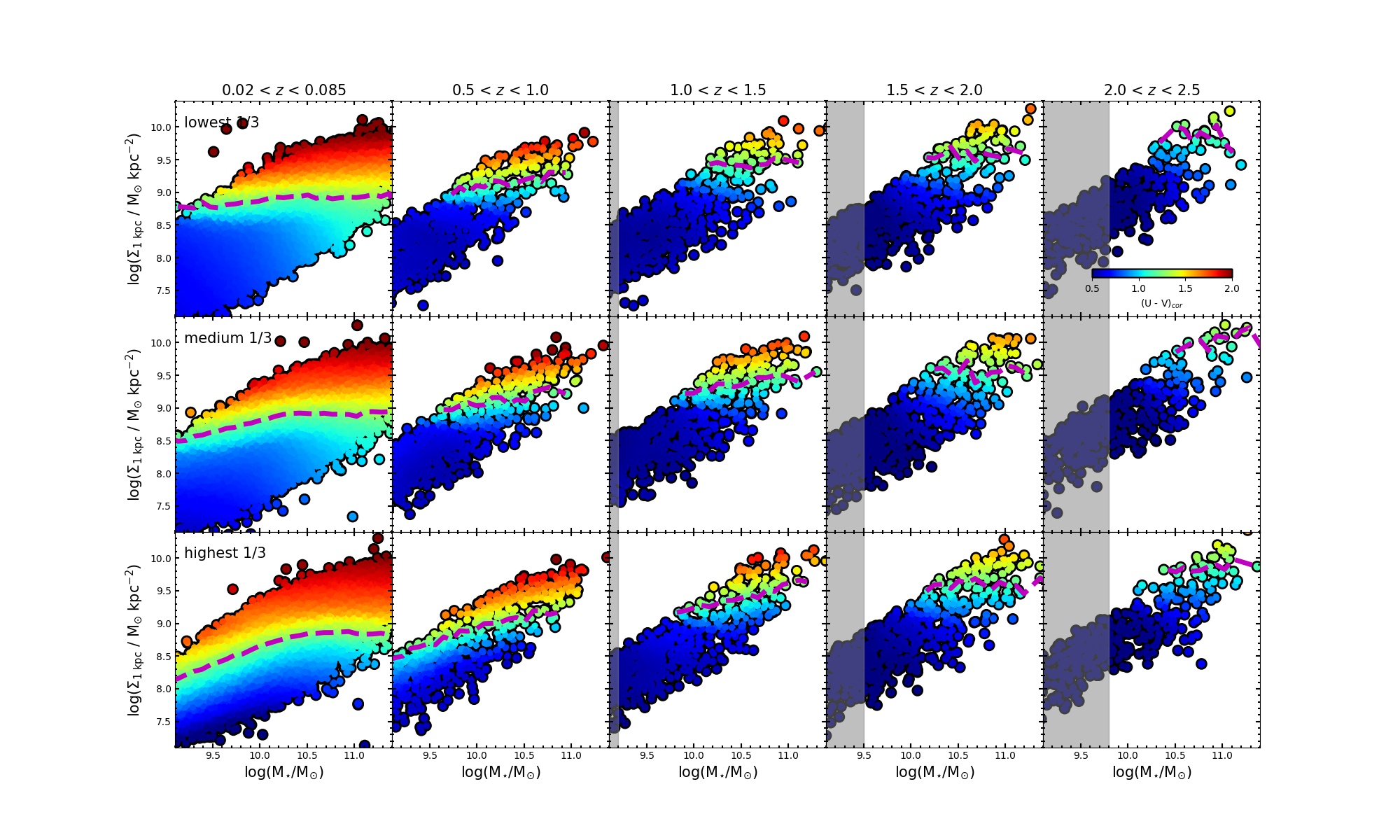}
		\caption{Central 1kpc surface mass density $\Sigma_{1\ kpc}$ as
		a function of stellar mass at five redshift bins (different 
		columns) and three environment bins (different rows), color-coded
		by the dust-corrected rest-frame (U - V) color. For the SDSS 
		spectroscopic sample, the data at $0.02 < z < 0.085$ has been 
		$V_{max}$-weighted; for the ZFOURGE photometric sample at $z 
		> 0.5$, the gray shaded region marks the range of stellar mass 
		which is incomplete.  The data in each bin has been LOESS 
		smoothed. The magenta dahsed lines denote the transitional 
		$\Sigma_{1\ kpc}$ at (U - V)$_{cor}$ $\sim$ 1.25.
		\label{fig:colorfig}}
\end{figure*}

Similar to the approaches in \citet{xu21}, we focus on the structural 
dependence on quenching by quantitatively sketching the trends in 
$\Sigma_{1\ kpc}$ at the transition from star-forming to passive 
populations, which is $(U - V)_{cor} \sim 1.25$ in this study, as discussed
in Sec \ref{subsec:cor} (also see Figure \ref{fig:uvc}). Figure 
\ref{fig:colorfig} shows the central 1kpc density $\Sigma_{1\ kpc}$ as a 
function of stellar mass M$_{\star}$ in three environmental bins and five 
redshift bins, color-coded by LOESS-smoothed, dust-extinction corrected 
color (U - V)$_{cor}$. In each bin, we select data at the transition that 
have $1.25-0.15 < (U - V)_{cor} < 1.25+0.15$ and $\Sigma^{crit}_{1\ kpc}$ 
is computed as the running median of their $\Sigma_{1\ kpc}$ as a function 
of stellar mass. We overplot the transitional $\Sigma_{1\ kpc}$ as the 
magenta dashed lines in Figure \ref{fig:colorfig} for reference. The 
(U - V)$_{cor}$ for SDSS galaxies in Figure \ref{fig:colorfig} is 0.3 dex 
lower than their original value as mentioned in Section \ref{subsec:cor}. 
We replot all galaxies with their original color in Figure 
\ref{fig:ms_sur_ori} in Appendix \ref{appen:ms_sur}. $\Sigma^{crit}_{1\ 
kpc}$ for SDSS galaxies remains unchanged under the shifting in color 
space, as expected.

Overall, there is strong redshift evolution in (U - V)$_{cor}$ color. At 
fixed stellar mass, the color at high redshifts is typically bluer than 
their low redshift counterparts. The critical $\Sigma_{1\ kpc}$ for massive
galaxies strongly evolves with redshift that the boundary appears higher at
high redshifts. At high redshift ($z > 1$), there is no significant 
difference in $\Sigma^{crit}_{1\ kpc}$ for massive galaxies in different 
environments; whereas the transitional line becomes environment-dependent
at low redshift ($z < 1$), in particular for low-mass galaxies.

\begin{figure*}
   \plotone{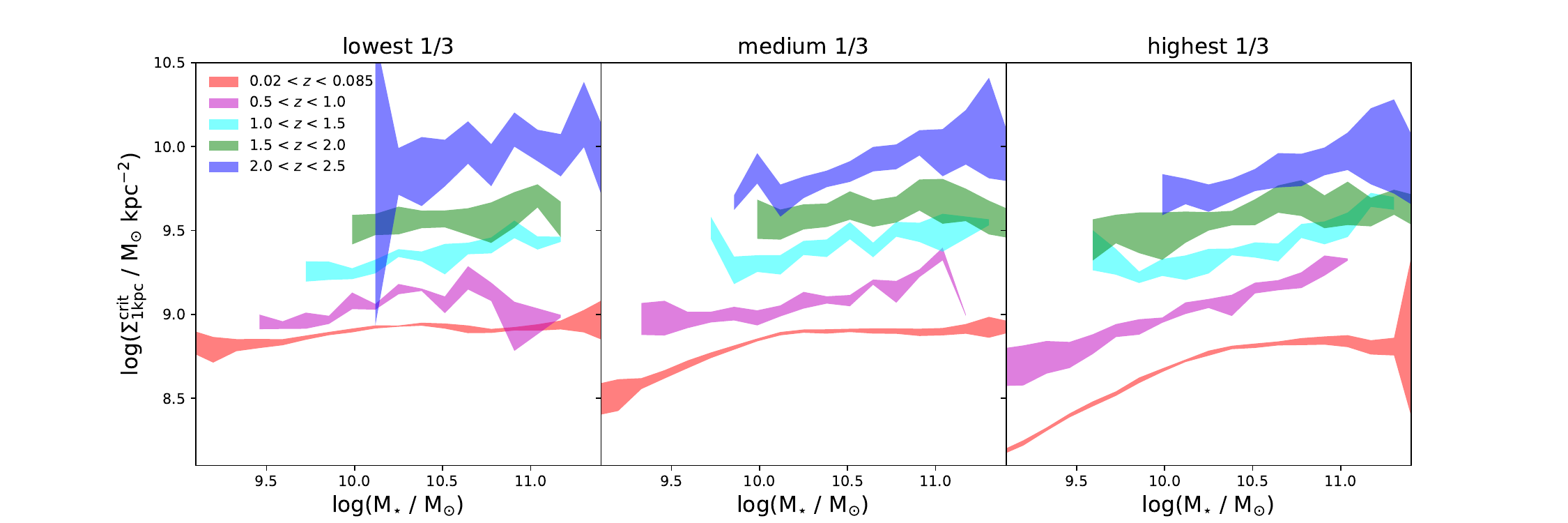}
        \caption{ $\Sigma^{crit}_{1\ kpc}$ as a function of stellar mass
        at five redshift bins and three environment bins. The color-shaded
        region marks the 1$\sigma$ error of the critical line which is
        computed by averaging over lines in 40 Monte-carlo simulated
        realizations in each redshift and environment bin.
        \label{fig:evo} }
\end{figure*}

We highlight these trends in Figure \ref{fig:evo} by plotting 
$\Sigma^{crit}_{1\ kpc}$ as a function of stellar mass in different 
environments at five redshift bins. The uncertainty of $\Sigma^{crit}_{1\ 
kpc}$ is estimated by Monte Carlo simulations. In each environment and 
redshift bin, we create 40 realizations of simulated data by perturbing 
various parameters. For nearby SDSS galaxies, we add a Gaussian random 
noise to the observed stellar mass and $\Sigma_{1\ kpc}$, with 1$\sigma$ 
uncertainty that is equal to their quoted errors. We also perturb the (U - 
V) color by adding noise to SDSS (u - g) and (g - r) colors with the quoted
1$\sigma$ error \citep{bla07}, and then propogating the errors 
using Equation \ref{eqn:color}. For ZFOURGE galaxies, we perturb their 
stellar mass, $\Sigma_{1\ kpc}$ and photometric redshift by adding
a Guassian random noise with their quoted 1$\sigma$ errors, we then rebin 
the simulated data based on their perturbed redshift and re-calculate the 
local overdensity in each realization, to account for the uncertainty in 
the photometric redshift. The colored shades in Figure \ref{fig:evo} are 
the running median and 1$\sigma$ dispersion in $\Sigma^{crit}_{1\ kpc}$.

The structural and environmental dependence on quenching and their 
evolution can be clearly depicted in Figure \ref{fig:evo}. At $z > 1$, 
no clear environmental dependence has been detected in $\Sigma^{crit}_{1\ 
kpc}$, and $\Sigma^{crit}_{1\ kpc}$ in all environments appears to be flat 
and exhibits only weak mass-dependency. The mass-dependence of 
$\Sigma^{crit}_{1\ kpc}$ in different environments becomes distinguishable
at $z < 1$: $\Sigma^{crit}_{1\ kpc}$ remains weakly mass-dependent in low
density, but rapidly increases with the stellar mass for low-mass galaxies
in dense environment, which is apparently due to the environmental effects.
In low density, $\Sigma^{crit}_{1\ kpc}$ only exhibits weakly 
mass-dependency and shows significant redshift evolution, which decreases 
by $\sim$ 1 dex from $\rm log(\Sigma^{crit}_{1\ kpc}) \sim 10$ at 
$2 < z < 2.5$ to $\rm log(\Sigma^{crit}_{1\ kpc}) \sim 8.8$ at $z \sim 0$; 
whereas in dense environment, $\Sigma^{crit}_{1\ kpc}$ evolves from being 
weakly mass-dependent at $z > 1$, to mildly mass-dependent at 
$0.5 < z < 1$, and eventually to strongly mass-dependent at $z \sim 0$.

We use another star-forming indicator -- sSFR to investigate the 
critical $\Sigma_{1\ kpc}$ on the plane of $\rm M_{\star}-\Sigma_{1\ kpc}$ 
in Figure \ref{fig:ms_sur_ssfr} in Appendix \ref{appen:ms_sur}. All trends 
in $\Sigma^{crit}_{1\ kpc}$ remain similar. Futhermore, to enable a 
complete assessment on the effects of environment, we also plot 
$\Sigma_{1\ kpc}$ as a function of the rank of log(1 + $\delta$) 
color-coded by (U - V)$_{cor}$ and sSFR for reference in Appendix 
\ref{appen:oden_sur}.

\section{Discussion and Summary}
\begin{figure*}
   \plotone{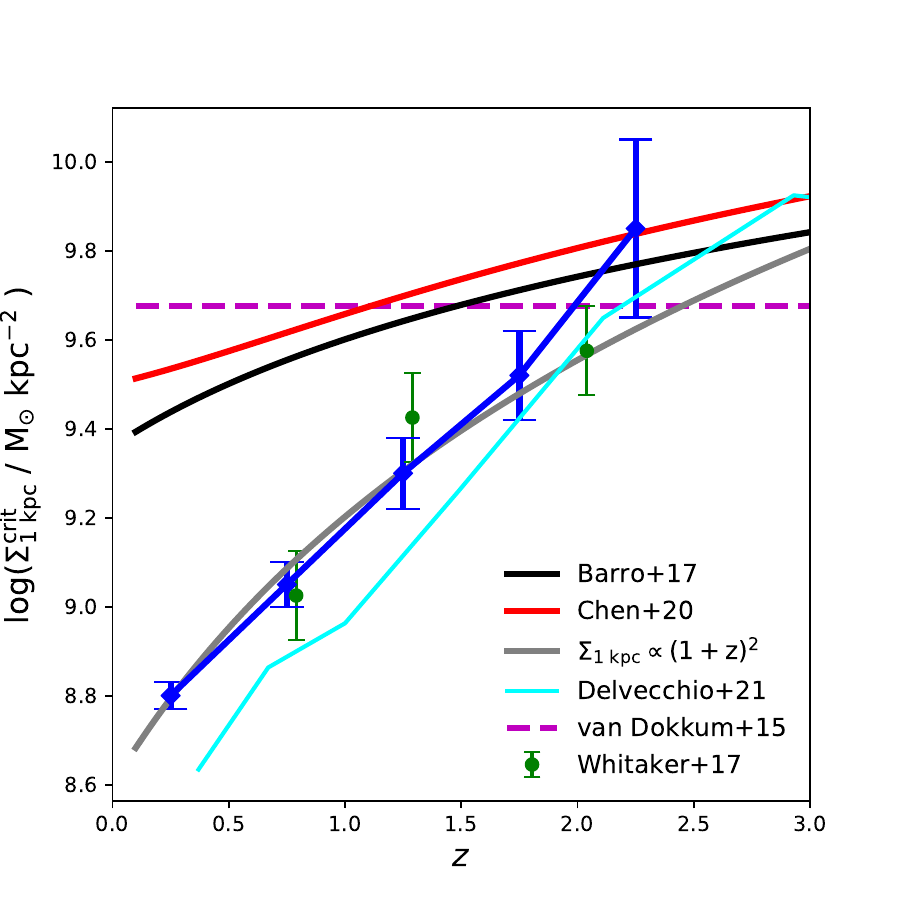}
		\caption{ $\Sigma^{crit}_{1\ kpc}$ as a function of redshift (blue
		diamonds). Green circles are the critical $\Sigma_{1\ kpc}$ from 
		\citet{whi17}, assuming an evolving limit in log(sSFR). Gray dashed
		line indicates the expected cosmological evolution in surface 
		density normalized at $z = 0.25$, $\Sigma_{1\ kpc} \propto(1+z)^2$.
		Red solid line denotes the redshift dependence of the normalization
		of the quenching boundary from \citet{che20}, normalized at $z = 
		2.25$. Black solid line marks the redshift evolution of the 
		zero-point of scaling relation of QGs from \citet{bar17}. Magenta 
		dashed line is the assumed constant threshold in velocity 
		dispersion above which galaxies quench at $1.5 < z < 3.0$ from 
		\citet{van15}. Cyan stars indicate the redshift evolution of $M_0$ 
		which correponds the ``bending" mass scale of the star-forming main
		sequence at each reshift. The data are from \citet{del21}, and 
		$M_0$ is converted to $\Sigma_{1\ kpc}$ using the scaling relations
		in \citet{bar17}.
		\label{fig:compar}    }
\end{figure*}

We use samples of nearby and distant galaxies from SDSS and ZFOURGE surveys
to explore the structural and environmental impacts on the galaxy quenching
and their redshift evolution, by studying the distribution of 
population-averaged color on the $M_{\star}-\Sigma_{1\ kpc}$ plane. In low 
density, we find that the critical $\Sigma_{1\ kpc}$ which separates the 
star-forming and passive populations, is only weakly mass-dependent. The 
presence of the weakly mass-dependent $\Sigma^{crit}_{1\ kpc}$ is first 
reported for central galaxies in local Universe \citep{xu21}, and our 
result confirms that the weakly mass-dependent $\Sigma^{crit}_{1\ kpc}$ 
appears ubiquitous at all redshifts, in particular for massive galaxies. We
use a different star-forming indicator (U - V)$_{cor}$ to compute 
$\Sigma^{crit}_{1\ kpc}$ in this study, comparing with the (NUV - r) color 
used in \citet{xu21}. Despite $\sim$ 0.3 dex offset in the computed 
$\Sigma^{crit}_{1\ kpc}$, the trend of weakly mass-dependency is not 
affected by the different choice of star-forming indicator. 
$\Sigma^{crit}_{1\ kpc}$ in low density exhibits strong redshift evolution 
with more than 1 dex decrement from $z = 2.5$ to $z = 0$ (blue data points 
in Figure \ref{fig:compar}). Meanwhile, $\Sigma^{crit}_{1\ kpc}$ in dense
environment are almost indistinguishable with their low-density 
counterparts at high redshifts, but becomes progressively mass-dependent
at $z < 1$, in particular for low-mass galaxies. As suggested in 
\citet{xu21}, the mass-dependent $\Sigma^{crit}_{1\ kpc}$ for nearby 
low-mass galaxies is due to the environmental effects. Our result confirms
that the ``bending" in $\Sigma^{crit}_{1\ kpc}$ for low-mass galaxies 
emerges at $z \sim 1$. This is not surprising, since it is well-established
that the environmental effects become predominant in quenching the low-mass
galaxies at $z < 1$ \citep{kaw17}. In addition, without an accurate
characterization of the environment at high redshifts, we are not able to 
identify various mass-dependency of $\Sigma^{crit}_{1\ kpc}$ across 
different environments. Therefore, the usage of ZFOURGE catalogs was 
initially motivated by the accuracy in measuring the photometric reshift, 
rather than the sample size. We notice that the photometric redshift in the
lastest released catalogs for CANDELS has been optimized with an 
uncertainty of $\sim\, 0.02 \times (1+z)$, by using statistics to correct 
the probability density functions (PDFs) of photometric redshift for biases
and errors \citep{kod23}. Data from larger surveys with similar accuracy in
photometric redshift such as CANDELS would certainly be valuable for our 
furture related studies.

\citet{whi17} use a sample of high-redshift galaxies from 3D-HST survey \citep{ske14} to study the population-averaged sSFR as a function of 
$\Sigma_{1\ kpc}$\footnote{In their work, a 3-D central 1kpc density $\rho_1$ is used, which
is $\rho_1 = \Sigma_{1\ kpc}$ + log(4/3).}. At each redshift, they derived 
a similar critical $\Sigma_{1\ kpc}$ at the transition from star-forming to
passive populations, which is defined by the population-averaged sSFR (e.g.
they explored both constant and evolving criteria in sSFR). They also find 
a strong evolution in $\Sigma^{crit}_{1\ kpc}$ (see green data points in 
Figure \ref{fig:compar}) which is well consistent with our result. 
\citet{van15} reported a threshold in central velocity dispersion of 225 
km s$^{-1}$, based on the analysis of a sample of compact star-forming 
galaxies at $1.5 < z < 3.0$ with a median size of $R_e = 1.8$ kpc. This 
quenching threshold increases to 234 km s$^{-1}$ after normalizing the 
velocity dispersion to $R_e = 1$ kpc following \citet{cap06}. This 
threshold in velocity dispersion is converted to a threshold in 
$log(\Sigma_{1\ kpc}) \sim 9.7$ (see magenta dashed line in Figure 
\ref{fig:compar}) using the correlation between the $\Sigma_{1\ kpc}$ and 
velocity dispersion \citep{fan13}, which also roughly agrees with our 
mass-averaged measurement of $\Sigma^{crit}_{1\ kpc}$ at $1.5 < z < 3.0$. 
The new finding in our study is that we explicitly investigate the stellar
mass and environmental dependences of $\Sigma^{crit}_{1\ kpc}$, which would
place additional constraints on the underlying quenching mechanisms.

On the other hand, \citet{bar17} use a sample of galaxies at 
$0.5 < z < 3.0$ from CANDELS survey to study $\Sigma_{1\ kpc}$ as a 
function of stellar mass $M_{\star}$ for SFGs and QGs, respectively. For 
QGs, they find a very tight $M_{\star}$-$\Sigma_{1\ kpc}$ scaling relation 
at all redshifts, and the typical scatter of the scaling relation is only 
0.14 dex. A similar scaling relation is also identified for SFGs, with a 
slightly larger scatter of $\sim$ 0.25 dex. They find that the slope of both scaling relations barely changes with the redshift ($\sim$ 0.9 for SFGs
and $\sim$ 0.66 for QGs), and only detect a weak redshift evolution in the 
zero-point ($\sim (1+z)^{\alpha}$, where $\alpha = 0.7-0.8$). More 
recently, \citet{che20} proposed a similar weak evolution of normalization 
of scaling relations for both populations, which can be parameterized as 
$h(z)^{-0.74}$, where $h(z) = H(z)/H(0)$.

The significance of the redshift evolution in $\Sigma^{crit}_{1\ kpc}$ may 
rely on the definition of the transitional core density, which varies in 
different studies. In \citet{bar17,che20}, $\Sigma^{crit}_{1\ kpc}$ is 
defined as a boundary computed from the scaling relation of QGs, below 
which few quenched galaxies can be found. At the same time, as argued in 
\citet{bar17}, such boundary is a necessary but not sufficient condition 
for quenching since there are some compact SFGs (cSFGs) found above the 
boundary. This boundary is essential to characterize the structural 
properties of the passive population, and its slow evolution (see black and
red solid lines in Figure \ref{fig:compar}) likely reflects that a higher 
core density is a prerequisite to quenching and the core density is less 
affected by the minor mergers that contribute majority of the size growth 
in passive galaxies. Meanwhile, in \citep{whi17,xu21} and this work, 
$\Sigma^{crit}_{1\ kpc}$ is computed based on a criteria of 
population-averaged color or sSFR. Though served in a statistical sense, 
the transitional core density defined in this way implicitly encodes the 
information of relative abundance of star-forming and passive populations 
in the sense that the abundance of QGs should be higher than that of SFGs 
above the boundary. Therefore, the strong evolution in 
$\Sigma^{crit}_{1\ kpc}$ (see blue diamonds in Figure \ref{fig:compar}) 
may attribute to the strong evolution in the number density of star-forming
and passive populations (see Figure \ref{fig:uvc} and \ref{fig:hist_ssfr}),
combined with that $\Sigma_{1\ kpc}$ in QGs is typically higher than that 
of SFGs, at fixed stellar mass.

The weakly stellar mass-dependent $\Sigma^{crit}_{1\ kpc}$ implies that 
mass-quenching is more sensitive to the central core density than other 
global properties such as stellar mass. A natural candidate machanism is 
AGN feedback, since the BH mass is closely related to the central velocity
dispersion and the central mass density \citep{fan13,blu14,blu20,che20}. As
discussed in \citet{xu21}, the weakly mass-dependent
$\Sigma^{crit}_{1\ kpc}$ is qualitatively consistent with the AGN feedback
model prescription in IllustrisTNG \citep{ter20}, though the $M_{BH}^{crit}
 \sim 10^{8.2} M_{\odot}$ implemented in TNG is systematically higher than
their inferred value in the local universe.

However, it is still challenging to account for the strong evolution in the
population-averaged color or sSFR at fixed stellar mass and 
$\Sigma_{1\ kpc}$ (see Figure \ref{fig:colorfig} and 
\ref{fig:ms_sur_ssfr}), or equivalently the strong evolution in 
$\Sigma^{crit}_{1\ kpc}$ at fixed stellar mass, if the core density or the 
central balck hole is the only quenching engine, since the quenching 
timescale due to AGN feedback is expected to be much shorter than 
$t_{Hubble}$. We attempt to interpret such strong evolution by the 
following scenarios. First, In the scenario of AGN feedback quenching, the 
integrated power output of AGN is dictated by the black hole mass, and the 
shutdown of star formation initiates as the integrated energy from the 
black hole becomes larger than the gravitational binding energy of gas 
within the halo \citep{ter20,che20,pio22,blu23}. At a given stellar mass, 
the halo is on average more massive at higher redshift \citep{beh19,gir20},
therefore the star formation swarmed with more gas requires a higher level 
of integrated energy from black hole or a more massive black hole to 
quench.

Second, the quenching process is also affected by the thermodynamical 
status of gas. As argued in \citep{dek06}, at $z < 2$ in massive halos with
$M_h > M_{shock}$, the gas is inevitably heated by the virial shock, and it
ultimately becomes diluted and more vulnerable to the feedback processes 
including AGN feedback. Hence a lower level of integrated energy (or a less
massive black hole) may be required to quench the massive galaxies in halo
with $M_h > M_{shock}$ in the hot accretion regimes at $z < 2$. Meanwhile,
the gas at $z > 2$ is preferentially accreted in the form of cold stream
instead of shock-heated medium. Hence, a more massive halo along with a more
massive black hole are required to shock-heat the gas into ``puffy" medium, 
thus enable the quenching due to AGN feedback to proceed in such cold 
accretion regimes. Therefore, the strong evolution in 
$\Sigma^{crit}_{1\ kpc}$ is likely to be a manifestation of the evolution of
boundary between cold and hot medium, which is discussed in \citet{dek06}.

There are some observational evidence to support the increasing mass scale 
with redshift, above which the cold gas supply is significantly reduced. For
instance, some studies have argued that the ``bending" of the star-forming 
main sequence (SFMS) at high mass end is likely due to the diminished cold 
gas supply by halo shock-heating \citep{del21,dad22,pop23}. \citet{del21} 
use a parametric form to quantitatively describe the ``bending" of SFMS, and
find that the characteristic ``bending" mass $M_0$ has strong redshift 
evolution. We use the best-fitted $M_{\star}-\Sigma_{1\ kpc}$ scaling 
relations for SFGs to convert $M_0$ to $\Sigma_{1\ kpc}$ and plot them in 
Figure \ref{fig:compar} (cyan stars). Interestingly, the slope of redshift 
evolution of $M_0$ matches that of $\Sigma^{crit}_{1\ kpc}$ very well, which
lends support to the halo-heating scenario. However, such consistency 
remains only in a qualitative level, since halo-heating scenario predicts 
that the quenching should also strongly correlate with the halo or stellar 
mass, which is contradicting with the observed weakly mass-denpendent 
$\Sigma^{crit}_{1\ kpc}$. Moreover, galaxies that live in haloes with 
$M_h < M_{shock}$ are expected to be in cold accretion phase even at low 
redshifts, and quenching occurs in these galaxies cannot be interpreted by 
halo-heating solely. Therefore, it is likely that both AGN feedback and 
halo-heating are acting in concert to contribute the observed evolution in 
$\Sigma^{crit}_{1\ kpc}$.

Thirds, other alternative quenching mechanism may also become more effective
at low redshifts, such as angular momentum quenching \citep{pen20,ren20}. As
the disks grow with cosmic time, the average angular momentum of galaxies 
gradually increases. At low redshifts, the accreted gas spirals in with too 
high angular momentum to enable a fast radial gas inflow to feed the inner 
regions of galaxies. Instead, these infalling gas would settle onto an outer
ring that is stable against fragmentation and radial migration. The star 
formation in the inner regions of galaxies would eventually be terminated 
due to the reduced gas supply or strangulation. As a consequence, a lower 
level of integrated energy from black hole is required as the quenching 
power. 

Finally, we caution that the observed correlation between the quiescence 
and central mass density does not necessarily imply causality, and a higher
$\Sigma_{1\ kpc}$ is likely to be the consequence of the quenching 
processes. At this stage, neither are we able to distangle the 
forementioned scenarios, nor to ascertain the causal direction, given the 
current data. Future cold gas survey (both HI and HII) and detailed 
comparison with the predictions from numerical simulations will be crucial 
to unveil the underlying physics of this structural evolution accompanying 
the quenching.

\section{Acknowledgement}
The authors wish to thank the anonymous referee for constructive comments
that have improved the manuscript. This work is supported by the National 
Science Foundation of China (NSFC) Grant No. 12125301, 12192220, 12192222, and the science research grants from the China Manned Space Project with NO. CMS-CSST-2021-A07. 

\appendix
\section{Local Environmental Indicators for Galaxies at High Redshift} \label{appen:otest}
\begin{figure*}
		\epsscale{0.95}
		\plotone{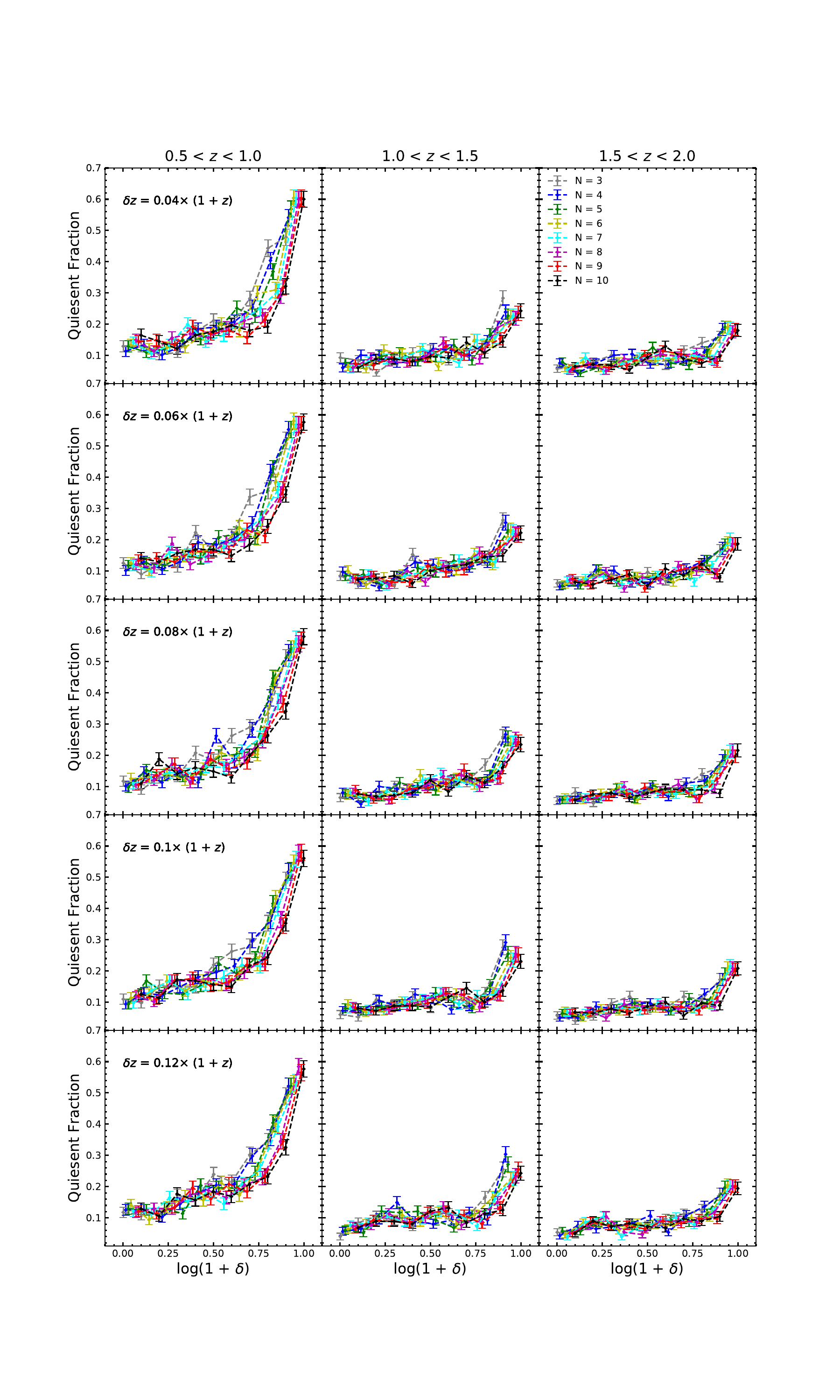}
        \caption{ Quiescent fraction as a function of local overdensity at 
        three redshift bins. Different row denotes different choice of 
		$\delta z$ in computing the overdensity. Eight choices of $N$ were 
		used in each panel to compute the overdensity, as indicated by their colors.\label{fig:qfrac_test}}
\end{figure*}

In this Appendix, we determine $N$ and $\delta z$ which are required to 
compute the local overdensity for the galaxies at high redshifts. We aim to
select $N$ and $\delta z$ upon which the constructed local overdensity
has the highest sensitivity to the the quiescence of galaxies. To achieve 
this, we study the quiescent fraction as a function of the rank of 
overdensity at three redshift bins: $0.5 < z < 1.0$, $1.0 < z < 1.5$ and 
$1.5 < z < 2.0$. In each redshift bin, we define 40 different local 
overdensities based on 8 choices of $N = 3,4,5,6,7,8,9,10$ and 5 choices of
$\delta z = 0.04,0.06,0.08,0.1,0.12\times (1+z)$. The quiescent galaxies 
are selected via UVJ diagram and the quiescent fraction of galaxies is 
computed in an interval of 0.1 in the rank of overdensity for each 
definition, as shown in Figure \ref{fig:qfrac_test}. 

\begin{figure*}
		\plotone{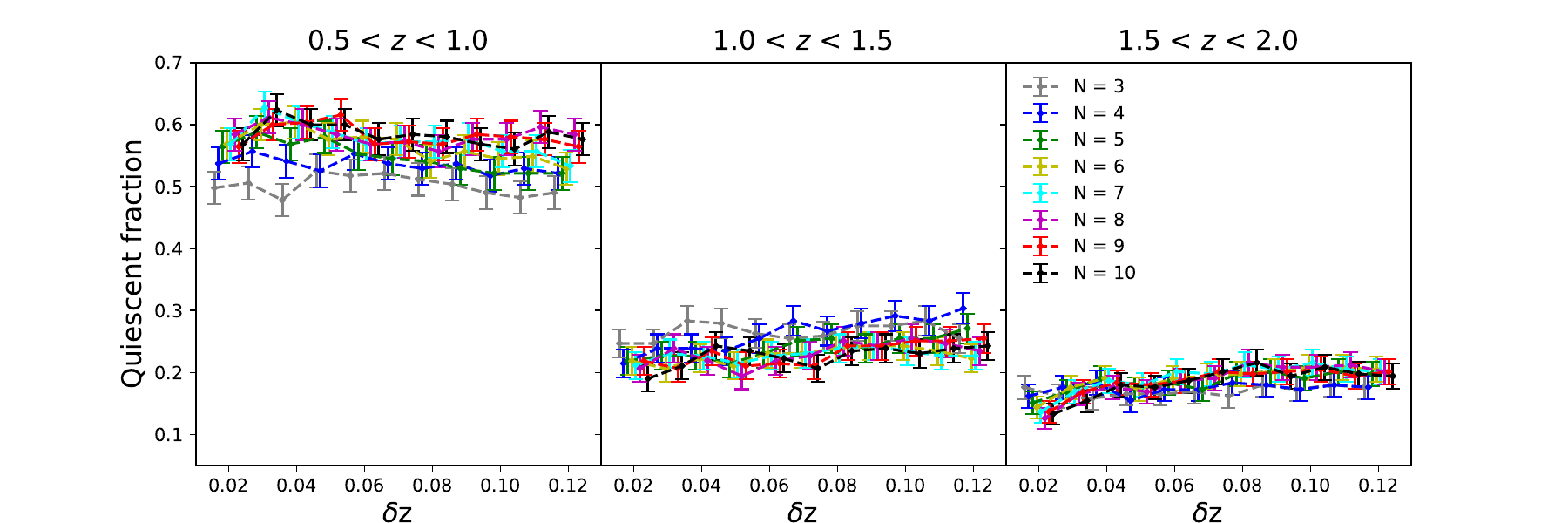}
		\caption{ Quiescent fraction in the densest bin as a function of 
		$\delta z$ at three redshift bins. Eight choices of $N$ were used to
		compute the  Quiescent fraction in the densest bin, as indicated by 
		their colors. \label{fig:otest}}
\end{figure*}

We define an indicator which is the quiescent fraction in the densest bin 
(e.g. highest 10\%), to quantitatively evaluate the sensitivity of 
quiescence to the local overdensity. We list the quiescent fraction in the 
densest bin as a function of $\delta z$ with multiple choices of $N$ in 
Figure \ref{fig:otest}. The best choice of $N$ and $\delta z$ appears to 
depend on the redshift, and there is no single choice of $N$ and $\delta z$ 
that prevails at all redshifts. For instance, at $0.5 < z < 1.0$, the 
quiescent fraction in densest bin is most sensitive to the overdensity with 
large $N$ ($N = 10$) and small $\delta z$ ($\delta z$ = 0.04$\times(1+z)$), 
while small $N$ ($N=4$) and large $\delta z$ ($\delta z$ = 
0.12$\times(1+z)$) appears a better choice at $1.0 < z < 1.5$. We then add 
up the quiescent fraction over all the three redshit bins for each $N$ and 
$\delta z$, and rank their performance. We find that the overdensity based 
on $N = 8$ and $\delta z = 0.08$ stands out to have the highest total 
quiescent fraction. Therefore, we use $N = 8$ and $\delta z = 0.08$ to 
compute the local overdensity of the galaxies at $0.5 < z < 2.5$ in this 
study.

\section{The correction of Dust Extinction} \label{appen:dust}
We evaluate the suitable $R_V$ used in the dust extinction law to correct
the (U - V) color for the effect of dust extinction in this Appedix. We 
first utilize the UVJ diagram the select SFGs and QGs at $0.5 < z < 2.5$. 
We follow the same selection criteria used in \citet{kaw17}, in which the 
rest-frame color satisfies:
  \begin{equation}
    \begin{aligned}
		  (U - V)_0 & > 1.2 \times (V - J)_0 + 0.2 \\
		  (U - V)_0 & > 1.3	\\
		  (V - J)_0 & < 1.6.		
    \end{aligned}		
  \end{equation}

\begin{figure*}
		\plottwo{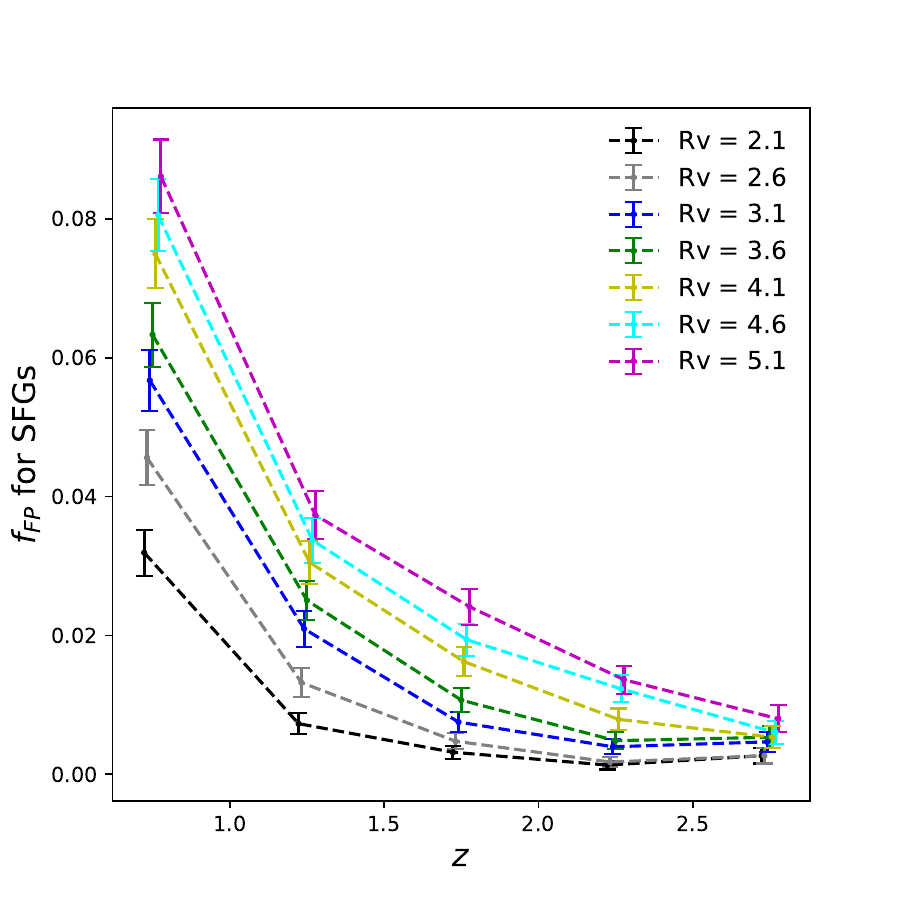}{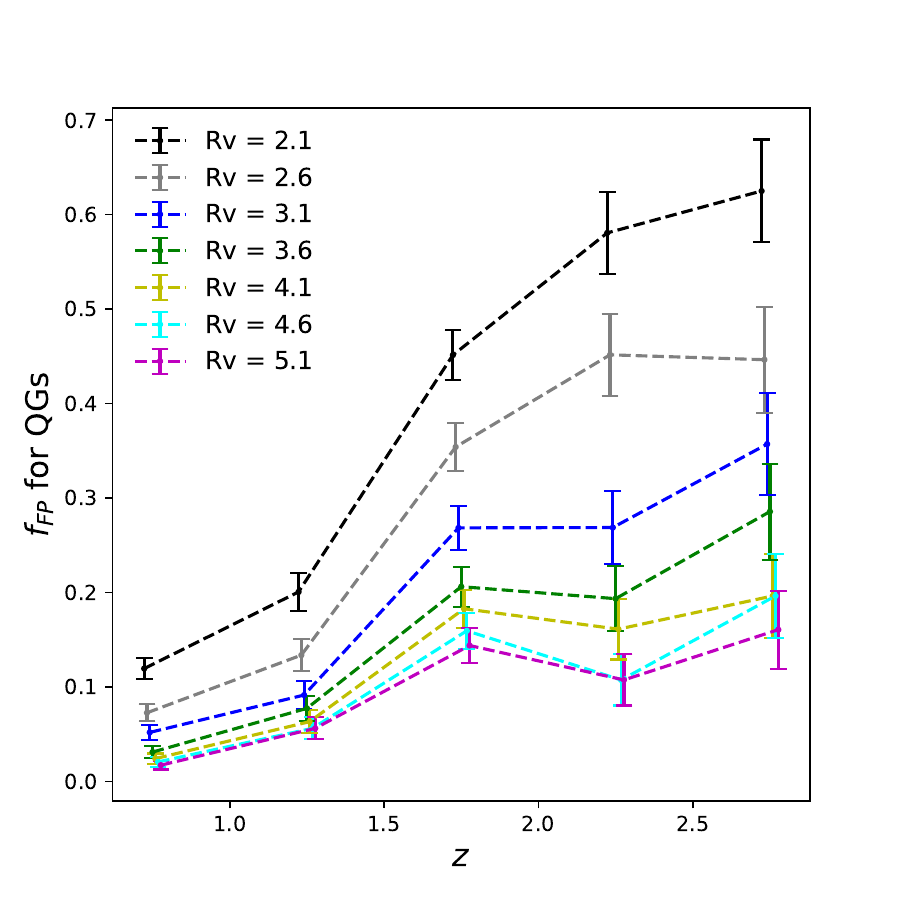}
		\caption{False positive rate for SFGs (QGs) as a function of 
		redshift in left (right) panel. Seven $R_V$ were used in computing
		the false positive rate, as indicated by their colors.
		\label{fig:fp}}
\end{figure*}

We then test 7 choices of $R_V = 2.1,2.6,3.1,3.6,4.1,4.6,5.1$. For each 
galaxy with a given value of $A_V$, we compute 7 corrected (U - V) color by 
assuming a Calzetti law and a choice of $R_V$. We find that the corrected 
color shows clear bimodality at almost all redshifts, and the color criteria
that seperates the SFGs and QGs is always $\sim 1.2$ regardless of choice of
$R_V$. Therefore, for each $R_V$, we use (U - V)$_{cor} \sim 1.2$ as color 
criteria to classify SFGs and QGs. Each classification was compared 
with the one based on UVJ diagram and our aim is to select the 
classification with a suitable $R_V$ that best mimics the UVJ 
classification. We define the false positive rates for SFGs and QGs, 
respectively, to quantify the ``similarity" of two classification schemes: 
$f_{FP}$ for SFGs is defined as the ratio of number of SFGs$^{\rm UVJ}$ 
with (U - V)$_{cor} > 1.2$ to the number of SFGs$^{\rm UVJ}$, while 
$f_{FP}$ for QGs is defined as the ratio of number of QGs$^{\rm UVJ}$ with (U - V)$_{cor} < 1.2$ to the number of 
QGs$^{\rm UVJ}$. We plot $f_{FP}$ as a function of redshift in Figure 
\ref{fig:fp} for SFGs and QGs, respectively.

\begin{figure*}
     \plotone{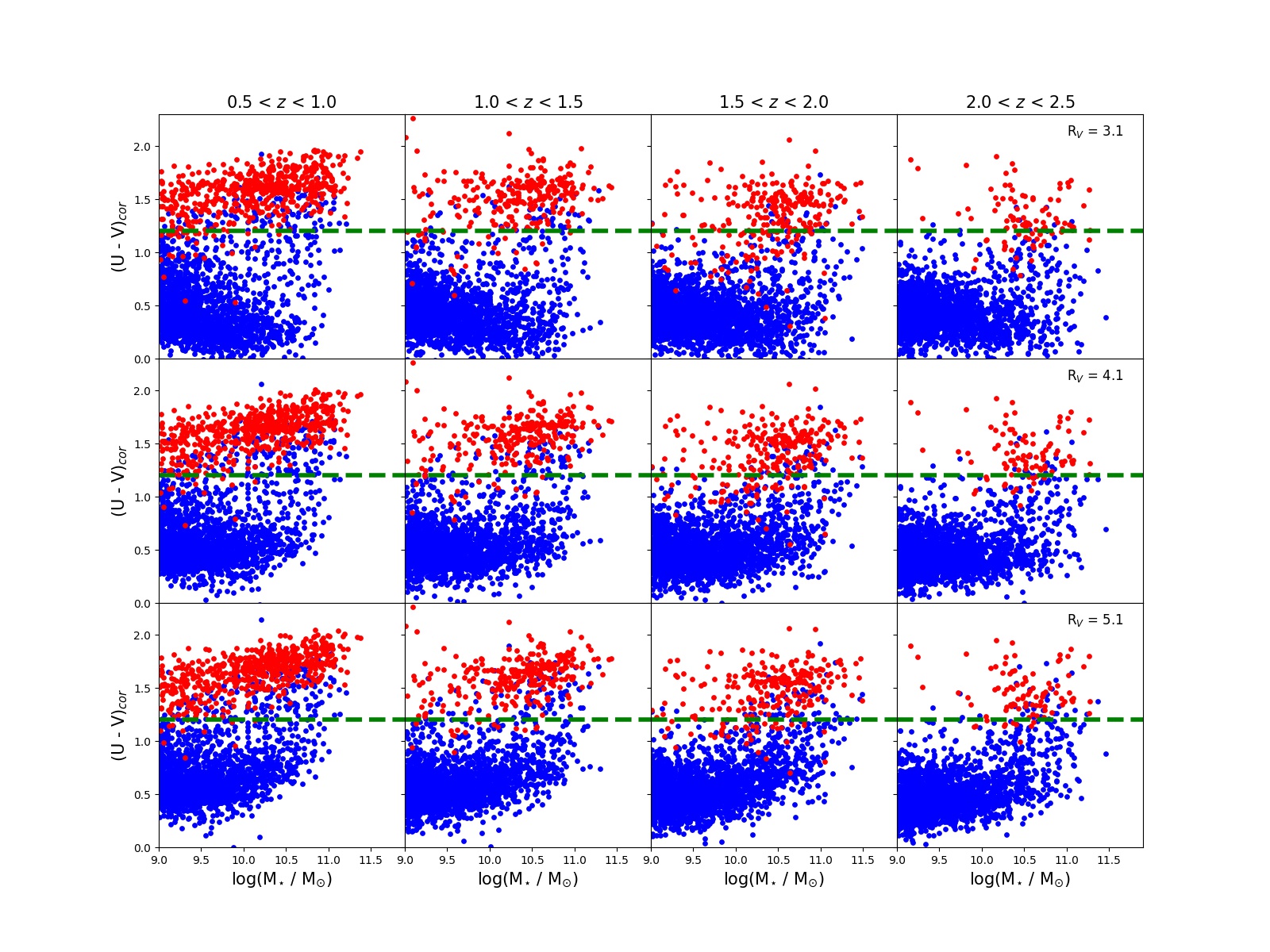}
       \caption{Dust-corrected rest-frame (U - V) color as a function of
		stellar mass at three redshift bins. Different row denotes different
		choice of $R_V$. Blue (red) data points are SFGs (QGs) identified 
		by UVJ diagram. Green dashed lines marks the color criteria that 
		seperates the star-forming and quenching in-process populations. 
		\label{fig:mstar_uvc}}
\end{figure*}

As Shown in Figure \ref{fig:fp}, $f_{FP}$ for SFGs (QGs) generally decreases
(increases) with the redshift. A higher $R_V$ tends to produce more(less) 
mis-classified SFGs(QGs). $f_{FP}$ for SFGs only has a mild redshift 
evolution, as $f_{FP}$ for SFGs is low on average (the highest $f_FP$ is 
$< 0.1$), and variation in $f_{FP}$ over different $R_V$ is also small ($< 
0.05$); while $R_V$ has an stronger impact on the evolution of $f_{FP}$ for
QGs, in particular at high redshifts. For instance, at $2.5 < z < 3.0$, 
$R_V = 2.1$ will result in $f_{FP} > 0.6$, which is much higher than 
$f_{FP} \sim 0.15$ with $R_V = 5.1$. We plot (U - V)$_{cor}$ as a function 
of stellar mass in four redshift bins with three different $R_V$ in Figure 
\ref{fig:mstar_uvc}. As shown, a QGs is more likely to be mis-classified as
a SFGs when using (U - V)$_{cor}$ with a lower $R_V$, in particular at 
high redshift. Therefore, we adopt a relatively high value of $R_V = 5.1$ 
in this study to correct the (U - V) color for galaxies at high redshifts. 

\section{Further explorations on stellar mass versus core density} \label{appen:ms_sur}

\subsection{$M_{\star}$ versus $\Sigma_{1\ kpc}$, color-coded by original 
(U - V)$_{cor}$ }
\begin{figure*}
   \plotone{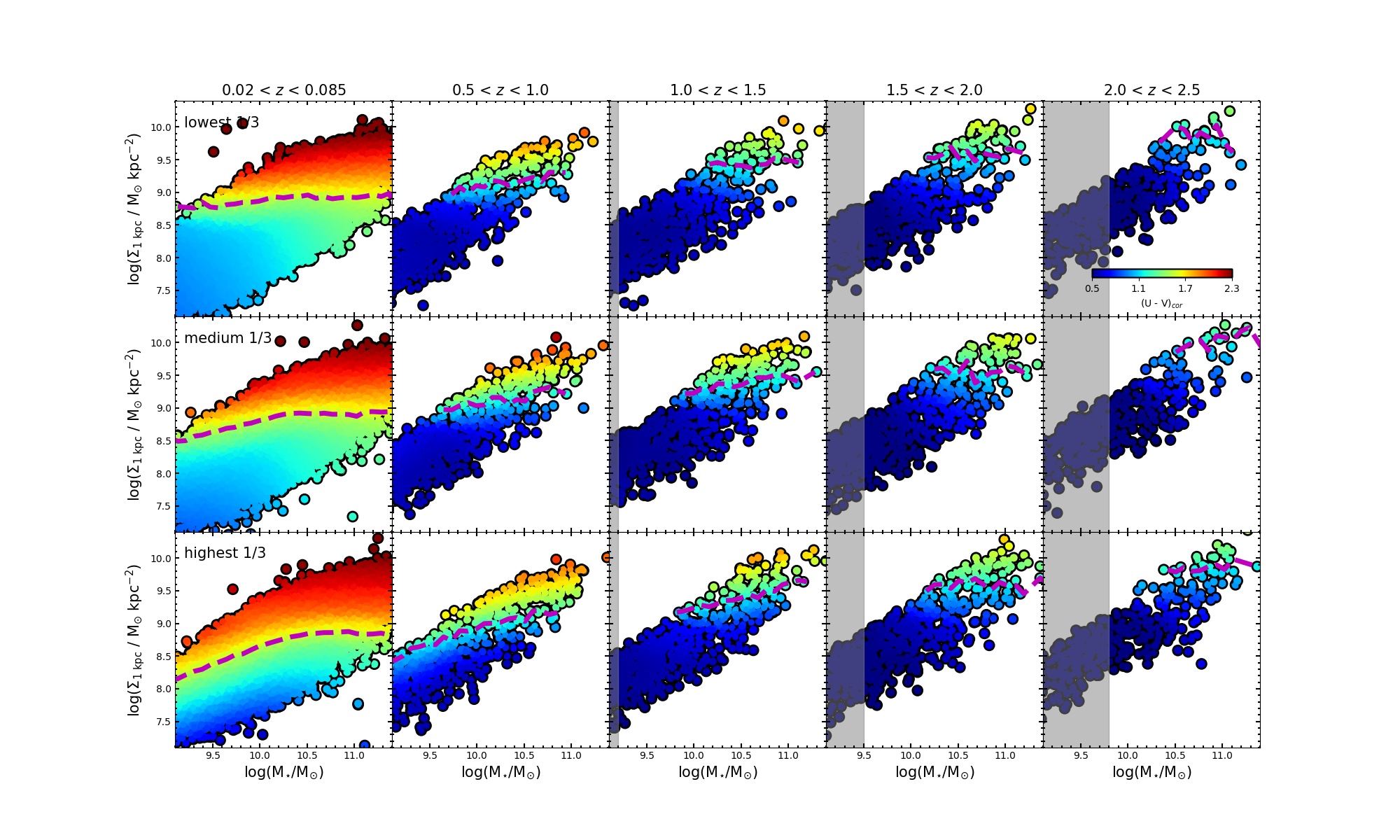}
		\caption{Similar to Figure \ref{fig:colorfig}. SDSS galaxies in 
		the first column are color-coded by their original (U - V)$_{cor}$ 
		without shifting. The magenta dashed lines denote the transitional 
		$\Sigma_{1\ kpc}$ at (U - V)$_{cor}$ $\sim$ 1.55, as marked by the 
		black dashed line in Figure \ref{fig:uvc}.}
        \label{fig:ms_sur_ori}
\end{figure*}

In Figure \ref{fig:uvc}, we shift the color distribution of SDSS galaxies 
towards the left by 0.3 dex, and use the shifted color to evaluate the 
critical $\Sigma_{1\ kpc}$  for SDSS galaxies in Figure \ref{fig:colorfig}. We 
emphasize that such shifting has no impact on the determination of 
$\Sigma^{crit}_{1\ kpc}$, but to narrow down the color range across all the
redshifts, and enhance the color contrast between the star-forming and 
quiescent populations for galaxies at high redshift. To avoid any confusion
on this shifting, we plot $\Sigma_{1\ kpc}$ as a function of stellar mass, 
color-coded by their original color in Figure \ref{fig:ms_sur_ori}. We 
repeat the same precedures as in Section \ref{sec:result} to compute 
$\Sigma^{crit}_{1\ kpc}$, and found that they indeed remain the same as in 
Figure \ref{fig:colorfig}.

\subsection{$M_{\star}$ versus $\Sigma_{1\ kpc}$, color-coded by sSFR}
In this Appendix, we explore the distribution of sSFR on the plane of 
$\rm M_{\star}-\Sigma_{1\ kpc}$. For SDSS galaxies, we utilize the data 
from the X2 version of Galaxy Evolution Explorer (GALEX)-SDSS-Wide-field 
Infrared Survey Explorer (WISE) Legacy Catalogue\footnote{https://salims.pages.iu.edu/gswlc/} 
\citep[GSWLC-X2,][]{sal16,sal18}, which is a value-added catalogue for
SDSS galaxies at $0.01 < z < 0.3$ within GALEX All-sky Imaging survey 
footprint \citep{mar05}. SFRs are derived from SED fitting consisting of 
two GALEX UV bands, five SDSS optical and near-IR bands, and one mid 
infrared band (22 microns if available, otherwise 12 microns) from WISE 
\citep{wri10}. For ZFOURGE galaxies, SFRs are estimated by combining the
total infrared luminosity ($L_{IR}$ = $L_{8-1000 \mu m}$) of galaxies and 
the luminosity emitted in the UV ($L_{UV}$ at rest-frame 2800 $\AA$). The 
extracted 24 -- 160 $\mu$m photometries from \emph{Spitzer}/IRAC and MIPS 
images (in all fields) and \emph{Herschel}/PACS images (available for 
CDFS only) were used to fit a model spectral template to calculate the 
total IR luminosity. $L_{UV}+L_{IR}$ provides an estimate of the total 
bolometric luminosity, which can be converted to SFR by \citep{bel05}:

\begin{equation}
		SFR\,[M_{\odot}\,yr^{-1}] = 1.09 \times 10^{-10}\, (L_{IR} + 2.2L_{UV})
\end{equation}		
We refer the readers to Section 6 in \citet{str16} for more details.

\begin{figure*}
   \plotone{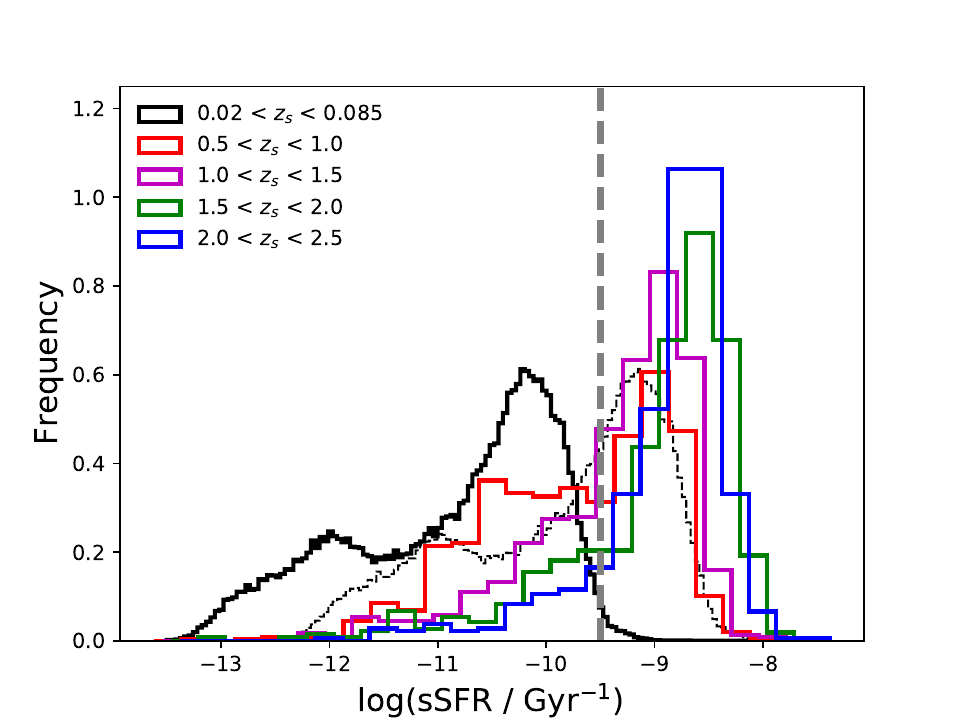}
        \caption{Distribution of specific SFR at five redshift bins.
        We shift the original disribution of sSFR for SDSS galaxies (black
        solid line) to the right by 1 dex (thin dashed line), to ease
        comparison of color distribution on the
        $\rm M_{\star}-\Sigma_{1\ kpc}$ plane. Gray dahsed line marks the
        critical sSFR that roughly separates the star-forming and quiescent
        populations at all redshifts,  which is
        $\rm log(sSFR / Gyr^{-1}) \sim -9.5$. }
        \label{fig:hist_ssfr}
\end{figure*}

We plot the distribution of sSFR for galaxies with $\rm 
log(M_{\star}/M_{\odot}) > 9.8$ at five redshift bins in Figure 
\ref{fig:hist_ssfr}. In general, sSFR also exhibits strong redshift 
evolution and galaxies have higher sSFR at higher redshifts. Similar to the
case of (U - V) color, a critical sSFR that separates the star-forming and 
passive populations is vital for the computation of 
$\Sigma^{crit}_{1\ kpc}$. As shown in Figure \ref{fig:hist_ssfr}, the 
star-forming population with $\rm log(sSFR / Gyr^{-1}) > -9.5$ dominates at
$z > 1$, and a noticable bump with $\rm log(sSFR / Gyr^{-1}) < -9.5$ 
emerges at $0.5 < z < 1$, which represents the increasing passive 
population. sSFR at $z \sim 0$ (SDSS galaxies) is much lower than their
high redshift counterparts, and shows clear bimodality. Following the same 
logic to ease comparison as in Figure \ref{fig:uvc}, we shift the 
distribution of sSFR by 1 dex towards the right in Figure \ref{fig:hist_ssfr} (thin dashed line). We refrain from sophisticatelly modeling the 
evolving criteria in sSFR as a function of redshift, since we only attempt to inspect if the trend in $\Sigma^{crit}_{1\ kpc}$ based on sSFR is 
consistent with that on (U - V)$_{cor}$. Instead, we adopt a simple 
criteria $\rm log(sSFR / Gyr^{-1}) \sim -9.5$ that roughly separates the 
two populations at all redshifts. Moreover, since the passive population in
this study not only consists of fully quenched galaxies, but also galaxies 
that are in the process of being quenched (e.g. Green Valley galaxies), the
adopted critical sSFR is slightly higher than the commonly quoted criteria
in literatures.

\begin{figure*}
   \plotone{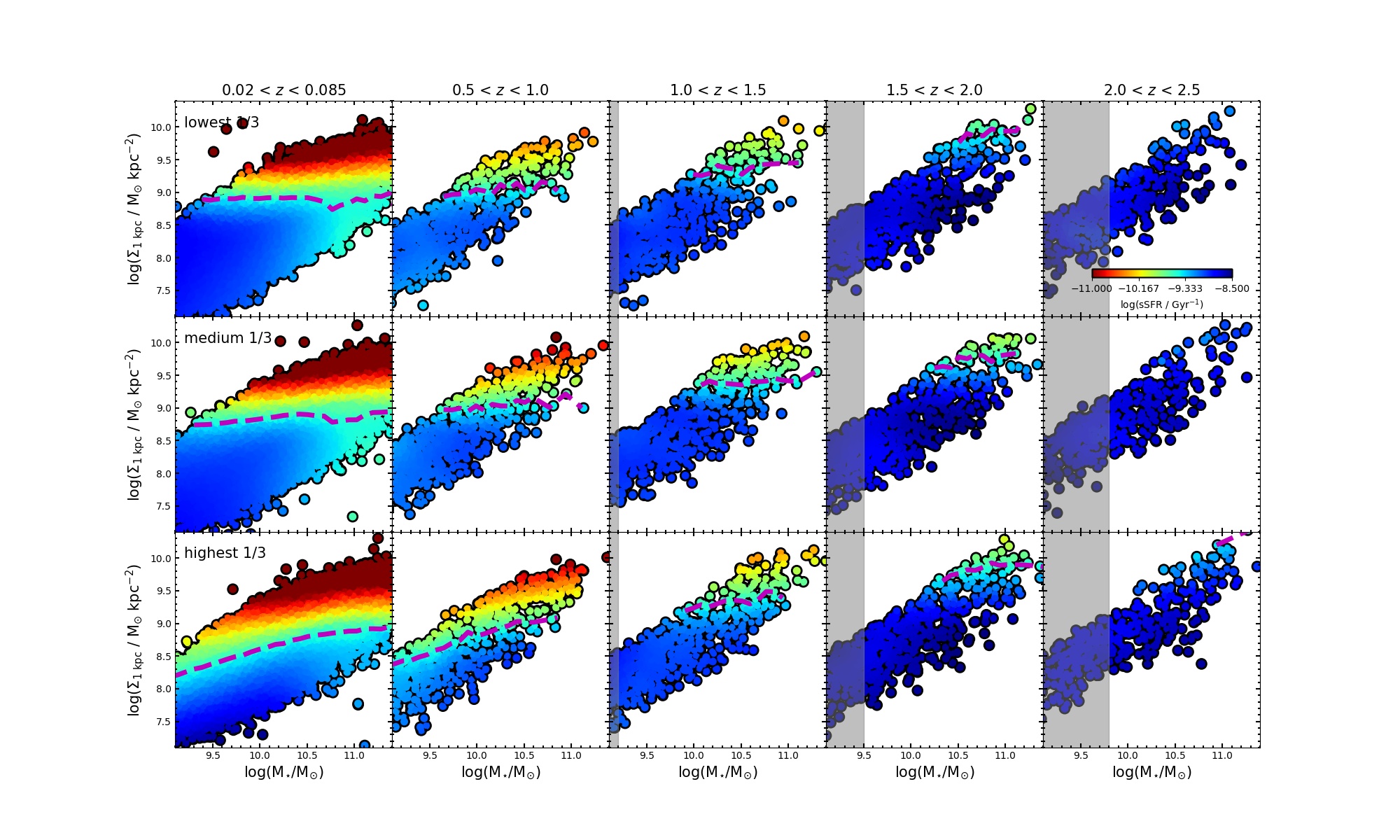}
		\caption{Similar to Figure \ref{fig:colorfig}. Galaxies are 
		color-coded by their sSFR. The magenta dashed lines denote the 
		transitional $\Sigma_{1\ kpc}$ at $\rm log(sSFR / Gyr^{-1}) \sim 
		-9.5$.}
		\label{fig:ms_sur_ssfr}	
\end{figure*}

We follow the similar procedures described in Section \ref{sec:result} to
assign galaxies into several redshift and environment bins, and plot 
$\Sigma_{1\ kpc}$ as a function of stellar mass, color-coded by their sSFR
in Figure \ref{fig:ms_sur_ssfr}. $\Sigma^{crit}_{1\ kpc}$ is computed as 
the running median of $\Sigma_{1\ kpc}$ that have 
$\rm -9.5 - 0.15 < log(sSFR / Gyr^{-1}) < -9.5 + 0.15$. We overplot 
$\Sigma^{crit}_{1\ kpc}$ as the magenta dashed lines in Figure 
\ref{fig:ms_sur_ssfr}. As shown, the trends in $\Sigma^{crit}_{1\ kpc}$ are
very similar with those base upon (U - V)$_{cor}$.

\section{Stellar mass versus overdensity} \label{appen:oden_sur}
\begin{figure*}
      \plotone{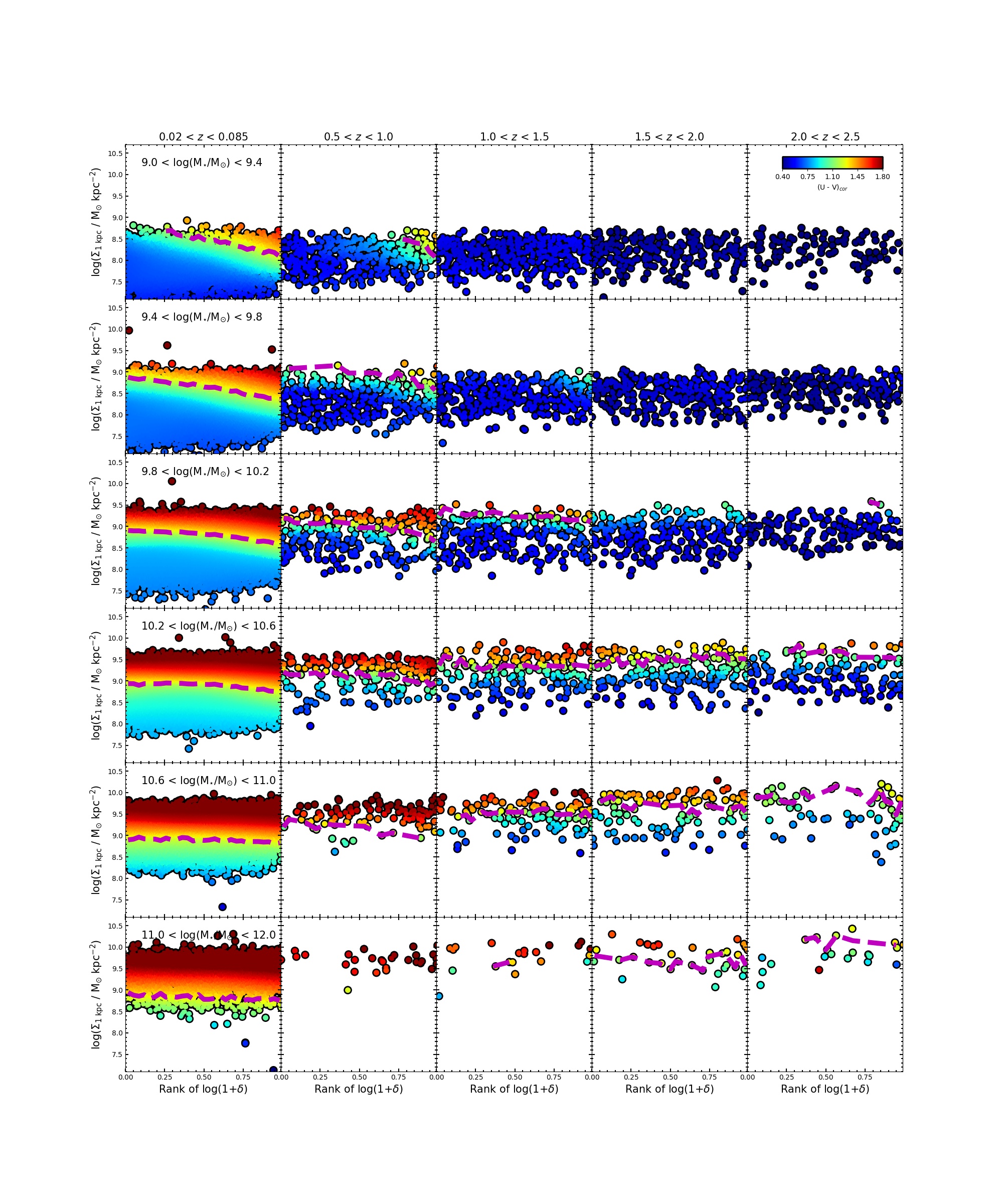}
      \caption{Central 1kpc surface mass density $\Sigma_{1\ kpc}$ as a 
		function of the rank of overdensity log(1+$\delta$) at five 
		redshift bins (different columns) and six stellar mass bins 
		(different rows), color-coded by the dust-corrected (U - V) color.
		the data at $0.02 < z < 0.085$ (SDSS galaxies) has been 
		$V_{max}$-weighted. The data in each bin has been LOESS smoothed.
		The magenta dahsed lines denote the transitional $\Sigma_{1\ kpc}$ 
		at (U - V)$_{cor}$ $\sim$ 1.25.}
        \label{fig:oden_sur}
\end{figure*}

In this Appendix, we investigate the relationship between the core density
and the overdensity at $0 < z < 2.5$, to gain complete insight of the 
effects of environment. We assign galaxies into five redshift bins to study
their redshift evolution. We divide the galaxies in each redshift bin into 
six stellar mass bins to study their stellar mass dependence. We perform a 
$V_{max}$-weighting correction for each SDSS galaxy to correct for the
incompleteness, inside a box of $0.2\times0.2$ dex$^2$ that centers on each
data point. The data in each bin has been LOESS smoothed. 
$\Sigma^{crit}_{1\ kpc}$ is computed based on the color criteria of 
(U - V)$_{cor}$ $\sim$ 1.25. We plot $\Sigma_{1\ kpc}$ as a function of the
rank of log(1 + $\delta$), color-coded by their (U - V)$_{cor}$ in Figure 
\ref{fig:oden_sur}. $\Sigma^{crit}_{1\ kpc}$ are overplotted as the magenta
dashed lines for reference.

For massive galaxies with $log(M_{\star}/M_{\odot}) > 10.2$, the color
is very strongly correlated with $\Sigma_{1\ kpc}$ and independent on the
environment, as indicated by the almost flat $\Sigma^{crit}_{1\ kpc}$. 
$\Sigma^{crit}_{1\ kpc}$ in all environments exhibits similar strong 
redshift evolution as depicted in the left panel of Figure \ref{fig:evo}. 
This is likely due to that large portion of massive galaxies are 
mass-quenched at high redshift, when the enviromental effects have yet to 
come into play. Quenching in these galaxies might occur in advance of their
infall into dense environments. 

For galaxies with intermediate and low mass, the environmental impact 
becomes progressively significant at $z < 1$, as dictated by the gradual 
tilt in $\Sigma^{crit}_{1\ kpc}$. At fixed stellar mass, 
$\Sigma^{crit}_{1\ kpc}$ appears inversely proportional to the overdensity.
For instance, in the lowest mass bin of $9.0 < log(M_{\star}/M_{\odot}) < 
9.4$ at $0.02 < z < 0.085$, $\Sigma^{crit}_{1\ kpc}$ in densest environment
is $\sim 0.7$ dex lower than that in sparse environment. Such inversely 
proportionality was first reported in \citet{xu21} for nearby galaxies, and
our study shows that it has already been in place at $z \sim 1$ (see second
column in Figure \ref{fig:oden_sur}). It could be qualitatively accounted
for by two facts, as suggested in \citet{xu21}: 1. there are more quenched
galaxies in dense environments at fixed stellar mass; 2. the color of 
galaxies is strongly correlated with their structure such as $\Sigma_{1\ 
kpc}$. Intriguingly, the environmental effects such as gas stripping is not
supposed to significantly alter the stellar structure of galaxies. An
almost vertical $\Sigma^{crit}_{1\ kpc}$ is anticipated if gas stripping is
the dominant effect of environment. Therefore, a tilted but non-vertical
$\Sigma^{crit}_{1\ kpc}$ would cast valuable contraints on the underlying
physical mechanisms of environmental quenching. After all, a strong 
correlation between the color and $\Sigma_{1\ kpc}$ at fixed stellar mass
favors those scenarios in which the structure of galaxies could be altered
by the environmental effects, such as tidal interaction or minor merger.

\begin{figure*}
      \plotone{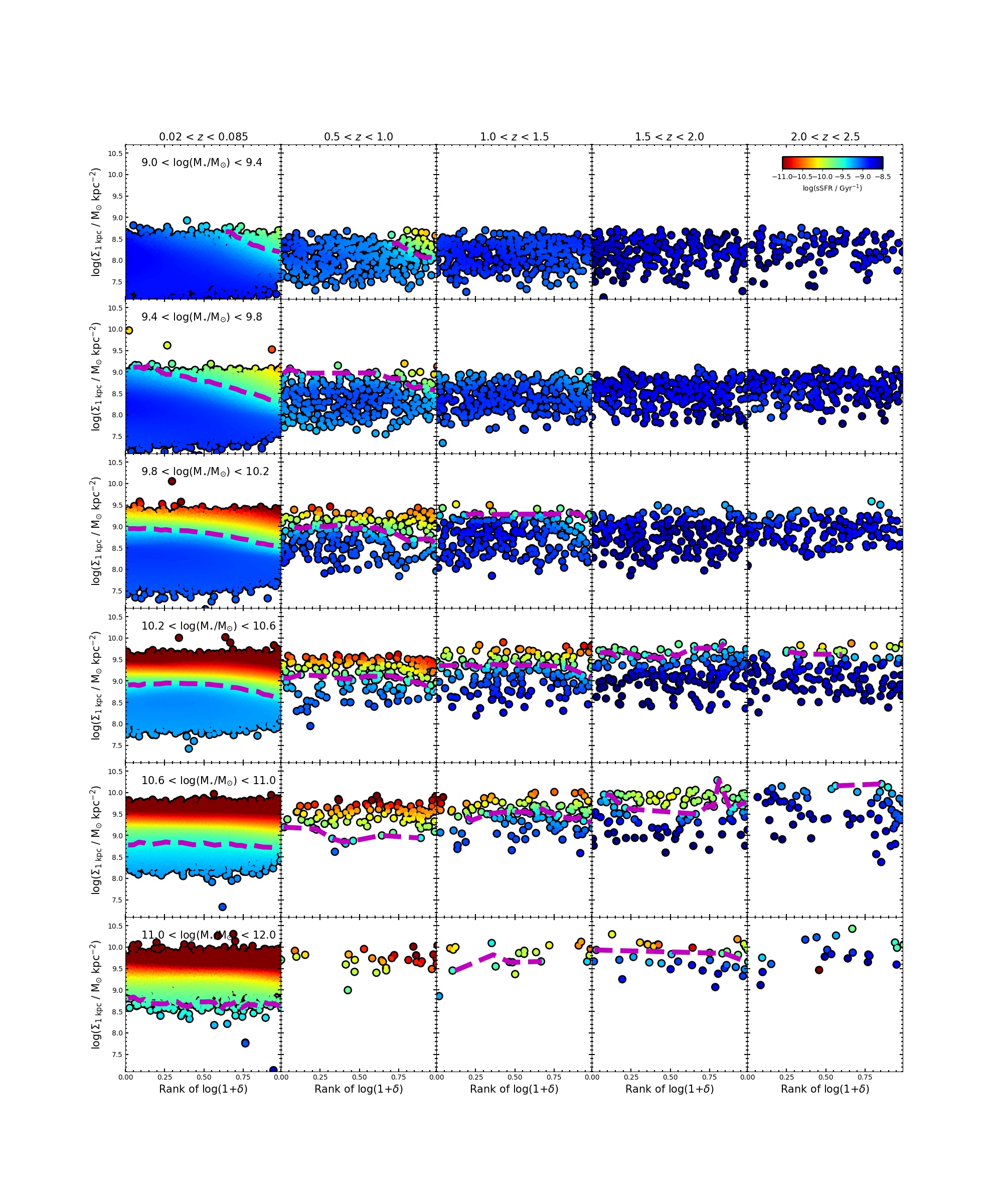}
      \caption{Similar to Figure \ref{fig:oden_sur}. Galaxies are 
		color-coded by their sSFR. The magenta dashed lines denote the
		transitional $\Sigma_{1\ kpc}$ at $\rm log(sSFR / Gyr^{-1}) \sim 
		-9.5$. }
        \label{fig:oden_sur_ssfr}
\end{figure*}

We also plot $\Sigma_{1\ kpc}$ as a function of rank of log(1 + $\delta$),
color-coded by their sSFR in Figure \ref{fig:oden_sur_ssfr} for reference. 
$\Sigma^{crit}_{1\ kpc}$ is computed based on 
$\rm log(sSFR / Gyr^{-1}) \sim -9.5$, as the magenta dashed lines denote.
The trends in $\Sigma^{crit}_{1\ kpc}$ remains similar as those in Figure 
\ref{fig:oden_sur}.

\bibliography{}
\bibliographystyle{apj}

\end{document}